\documentclass{emulateapj-rtx4}
\usepackage{amssymb, amsmath, epsfig,natbib}

\newcommand{\cc}{{\rm~ cm^{-3}}}

\newcommand{\Msun}{{M_{\odot}}}

\newcommand{\HI}{H{\sc ~i}}

\newcommand{\OVI}{O{\sc ~vi}}
\newcommand{\OVII}{O{\sc ~vii}}
\newcommand{\OVIII}{O{\sc ~viii}}

\newcommand{\CIII}{C{\sc ~iii}}

\newcommand{\SiIII}{Si{\sc ~iii}}
\newcommand{\SiIV}{Si{\sc ~iv}}

\newcommand{\NV}{N{\sc ~v}}
\newcommand{\NIII}{N{\sc ~iii}}

\begin{document}

\title{Implications of the large O{\sc ~vi} columns around low-redshift $L_*$ galaxies}
\author{Matthew McQuinn\altaffilmark{1,2} and Jessica K. Werk\altaffilmark{1}}

\altaffiltext{1} {Department of Astronomy, University of Washington, USA\\}
\altaffiltext{2} {Institute for Advanced Study, Princeton, NJ, USA\\}
\begin{abstract}
Observations reveal massive amounts of O{\sc ~vi} around star-forming $\sim L_*$ galaxies, with near unity covering fractions extending to the host halo's virial radius.  This O{\sc ~vi} absorption is typically kinematically centered upon photoionized gas, with line widths that are suprathermal and kinematically offset from the galaxy.   We discuss various scenarios and whether they could result in the observed phenomenology (cooling gas flows, boundary layers, shocks, virialized gas, photoionized clouds in thermal equilibrium).  If predominantly collisionally ionized, as we argue is most probable, the O{\sc ~vi} observations require that the circumgalactic medium (CGM) of $L_*$ galaxies holds nearly all the associated baryons within a virial radius ($\sim 10^{11}\Msun$) and hosts massive flows of cooling gas with $\approx 30 [n T/30{\rm \, cm^{-3} K}] \Msun$~yr$^{-1}$ through $3\times10^5$K, which must largely be disrupted and prevented from accreting onto the host galaxy.  Cooling and feedback energetics considerations require $10 {\rm \, cm^{-3} K} < \langle n T \rangle < 100  {\rm \, cm^{-3} K}$ for the warm and hot halo gases.  We argue that virialized gas, boundary layers, hot winds, and shocks are unlikely to directly account for the bulk of the O{\sc ~vi}. Furthermore, we show that there is a robust constraint on the number density of many of the photoionized $\approx 10^4$K absorption systems that yields upper bounds in the range $n< (0.1-3) \times 10^{-3} (Z/0.3)$cm$^{-3}$, where $Z$ is the metallicity, suggesting that the dominant pressure in some photoionized ``clouds'' is nonthermal.  This constraint, which requires minimal ionization modeling, is in accordance with the low densities inferred from more complex photoionization modeling.  The large amount of cooling gas that is inferred could re-form these clouds in a fraction of the halo dynamical time, as some arguments require, and it requires much of the feedback energy available from supernovae and stellar winds to be dissipated in the CGM. 
\end{abstract}


\section{introduction}

A rich phenomenology has emerged over the last several years for gas in the circumgalactic medium (CGM) of $\sim L_*$ galaxies, driven largely by a statistical sample of CGM absorption systems  observed with the Cosmic Origins Spectrograph (COS) on the Hubble Space Telescope \citep{green12}.   Perhaps the biggest surprise of the COS observations is the ubiquity of \OVI\ in absorption \citep{tumlinson11}.  Almost all sightlines that penetrate within one virial radius from a star-forming galaxy show columns of $N_{\rm HI} \sim 10^{14.5}{\rm~ cm}^{-2}$ \citep{2011ApJ...740...91P, tumlinson11,johnson15}, which translates to masses in \OVI-tracing gas of 
\begin{equation}
M                = 8\times 10^{9} \Msun  \left(\frac{0.1}{f_{\rm OVI} Z}\right)
                   \left( \frac{ N_{\rm OVI}}{10^{14.5}{\rm cm}^{-2}} \right) \left(\frac{R}{200 {\rm kpc}}\right)^2, 
                   \label{eqn:intro}
\end{equation}
where $N_{\rm OVI}$ is the average \OVI\ column, $R$ its effective radial extent, $Z$ the metallicity of oxygen in solar units, and $f_{\rm OVI}$ the ionic fraction in \OVI\ (which is $<0.2$ in collisional equilibrium).  Thus, \OVI\ gas traces a significant fraction of the total baryonic mass of these $10^{12}\Msun$ halos, especially since we expect $f_{\rm OVI} Z < 0.1$.  Additionally, in contrast to star-forming galaxies, in quiescent galaxies, \citet{tumlinson11} reported \OVI\ detections in only $4$ of $12$ cases, suggesting a link between \OVI\ and ongoing star formation.   

COS observations also probe lower ionization metal absorption systems that likely originate from $\sim 10^4$K photoionized regions.  These systems are seen in more than half of sightlines that lie within $\sim 100\,$kpc of the galaxy in projection \citep{werk13, stocke13}, and they often appear at coincident velocities as \OVI\ absorbers \citep{werk16}.  Like with the \OVI\ absorbers, they tend to have suprathermal line widths and velocity offsets from their host galaxies of $\lesssim 100~$km~s$^{-1}$, although unlike the \OVI\ they present similar properties in both star-forming and quiescent galaxies.   Photoionization models of these absorbers suggest that the lower ionization systems trace gaseous reservoirs of $\sim10^{10-11}\Msun$, $\sim 1-10\%$ of the halo masses \citep{stocke13, werk14, stern16, prochaska17}.

Complementing these extragalactic observations, the CGM of the Milky Way itself provides a suite of constraints on the properties of intrahalo gas.  The volume-filling phase has been constrained through the confinement and stripping of the Magellanic stream and of high-velocity clouds \citep{stanimirovic02, fox05}, Large Magellanic Cloud pulsar dispersion measurements \citep{2001ApJ...553..367C, 2006ApJ...649..235M}, and detections of \OVII\ and \OVIII\ in absorption and emission \citep{gupta12, miller15}.  These diverse methods all find densities of $n \sim 10^{-4}$cm$^{-3}$ for the $\sim 10^6$K volume-filling halo gas, and additional modeling suggests that this gas extends beyond $100~$kpc from the Galaxy with a total mass of $>10^{10}\Msun$ \citep{gupta12,fang13, miller15, faerman16, 2015ApJ...815...77S}.  Furthermore, observations of both neutral and ionized high-velocity clouds constrain the mass flow rates and distances of these cooler, largely infalling cloud-like structures, indicating mass accretion rates onto the disk of $\sim1-10~\Msun$yr$^{-1}$ and $\sim10~$kpc distances \citep{1997ARA&A..35..217W, 2011Sci...334..955L, richter17}.  The statistics of high-velocity clouds absorption systems and extragalactic absorbers around $L_*$ galaxies appear consistent when accounting for selection effects \citep{2015ApJ...807..103Z}. 

Significant work has been devoted to modeling the CGM, often building off of the more developed understanding for gas in galaxy groups and clusters \citep{sharma12b, 2017MNRAS.466..677G}.  CGM models have had the most success at reproducing the $n \sim 10^{-4}\,$cm$^{-3}$ Galactic corona \citep[e.g.][]{maller04, mccourt12, feldmann13}. The cooler CGM gas has proven more difficult to simulate.  Most previous cosmological simulations underpredict by factors of $3-30$ the amount of \OVI\ and of low ion absorption around $L_*$ star-forming galaxies \citep{hummels13, suresh15, ford15, oppenheimer16, fielding16}.  By tuning feedback recipes and incorporating additional physics (such as cosmic ray feedback), some simulations have achieved better agreement \citep{stinson12, salem16}.  
  
The goal of realistic simulations of the CGM is ambitious.  Such simulations must capture gaseous outflows from galaxies.  They may also need to resolve cloud-shredding instabilities as well as to capture the magnetic fields and conductive heat transport that, for one, play a role in shaping these instabilities.  Furthermore, non-equilibrium chemistry and the uncertain spectrum of the ambient radiation field can affect the CGM's ionization states and cooling times \citep{2007ApJS..168..213G, oppenheimer13, 2009MNRAS.393...99W, cantalupo10}.  Rather than simulate the full CGM, a complementary numerical approach is to simulate a specific hydrodynamic process that may be occurring in the CGM, such as turbulent boundary layers or thermal instability \citep[e.g.][]{kwak10, mccourt15}.  Studies in this vein inevitably encounter difficulty relating their calculations to the global picture. 

In light of these challenges, this paper takes another approach, in the spirit of \citet{mo96}, \citet{heckman02}, and \citet{maller04}, consisting of back-of-the-envelope analytics with empirical inputs.   We find that this approach, faced with the extensive set of observational constraints, supports certain CGM hypotheses over others.  In addition, this approach motivates the microphysical processes that are worthy of further numerical exploration, and it constrains the physics that must be captured in global CGM simulations.

This paper is organized as follows. Section~\ref{sec:CGMproperties} details the observed CGM properties that we find to be most constraining for CGM models, with a particular focus on \OVI\ absorption.  Section~\ref{sec:simple} presents simple calculations that are germane to many hypotheses for what creates the CGM \OVI, and, building off these calculations, Section~\ref{sec:hypotheses} discusses in more detail the viability of the different hypotheses. We turn to the photoionized (lower-ionization) metal absorption systems in Section~\ref{sec:HI}, presenting simple estimates for cloud sizes and masses.  
  We offer concluding remarks in Section~\ref{sec:conclusions}.  We elaborate on select aspects of our calculations with a series of appendices: Appendix~\ref{sec:kinematics} discusses relevant kinematical and hydrodynamical disruption timescales,  Appendix~\ref{sec:proximity} argues that our results are robust to uncertainties in both metagalactic and local sources of radiation, and Appendix~\ref{sec:helpful} serves as a reference for the cooling times and metal ionization fractions of gases at various densities and temperatures.

Let us address some parameter and notational choices that appear throughout this paper.  Many of our calculations involve the abundance of oxygen; we take the solar oxygen to hydrogen abundance ratio to be $[f_{\rm O}]_\odot = 5\times 10^{-4}$ \citep{2009ARA&A..47..481A}.  We will often phrase calculations in terms of pressures using the notation $P \equiv n T$, where $n$ is the electron density (which to our level of approximation is also the hydrogen density) and $T$ is the temperature.  We will also use the shorthand $T_4$, $n_{-4}$, $P_{2}$, and $R_{200{\rm k}}$: $T_4$ is temperature in $10^4$K, $n_{-4}$ is the number density in units of $10^{-4}{\rm~cm}^{-3}$, $P_{2}$ is $n T$ in units of $10^2\,$K~cm$^{-3}$, and $R_{200{\rm k}}$ is a distance in units of $200\;$kpc (an odd notation that attempts to avoid confusion with the standard nomenclature for the dark matter halo radius).  We will remind the reader of these notations throughout.  Although none of our calculations rely heavily on the background cosmology, where required we assume a flat $\Lambda$CDM model with $\Omega_m = 0.3$, $h=0.7$, and $Y_{\rm He} = 0.25$.  
 
 Lastly, this paper makes regular use of the ionic abundance tables of \citet{oppenheimer13}\footnote{\url{http://noneq.strw.leidenuniv.nl}}  and the COS-Halos absorption line catalogues of \citet{tumlinsonCOS} and \citet{werk13}.  When plotting the COS-Halos data, we generally use only lines that are unblended and unsaturated (see \citealt{werk13}). 

\section{Empirical CGM properties}
\label{sec:CGMproperties}

\begin{figure*}
\begin{center}
\epsfig{file=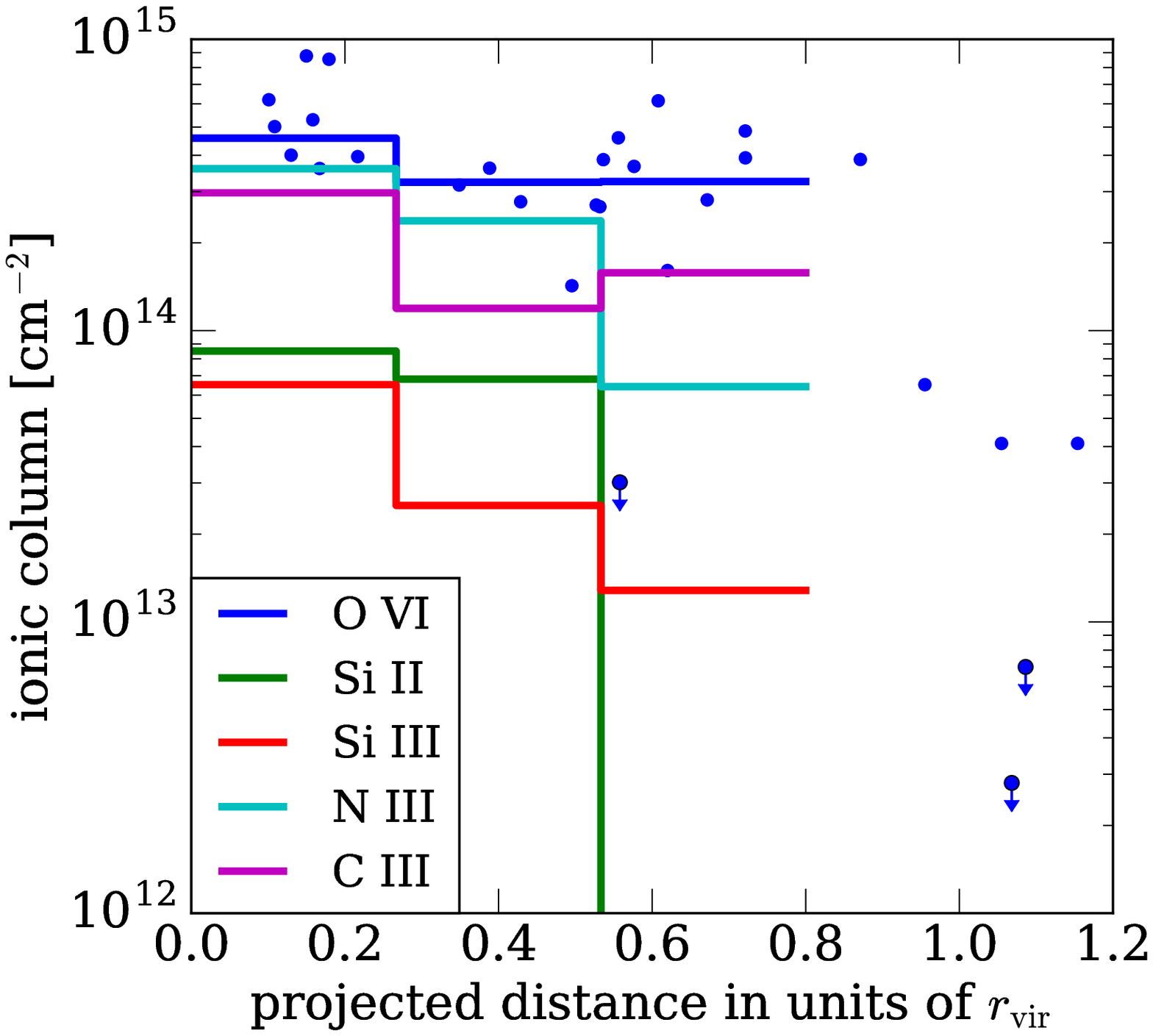, width=5.8cm} 
\epsfig{file=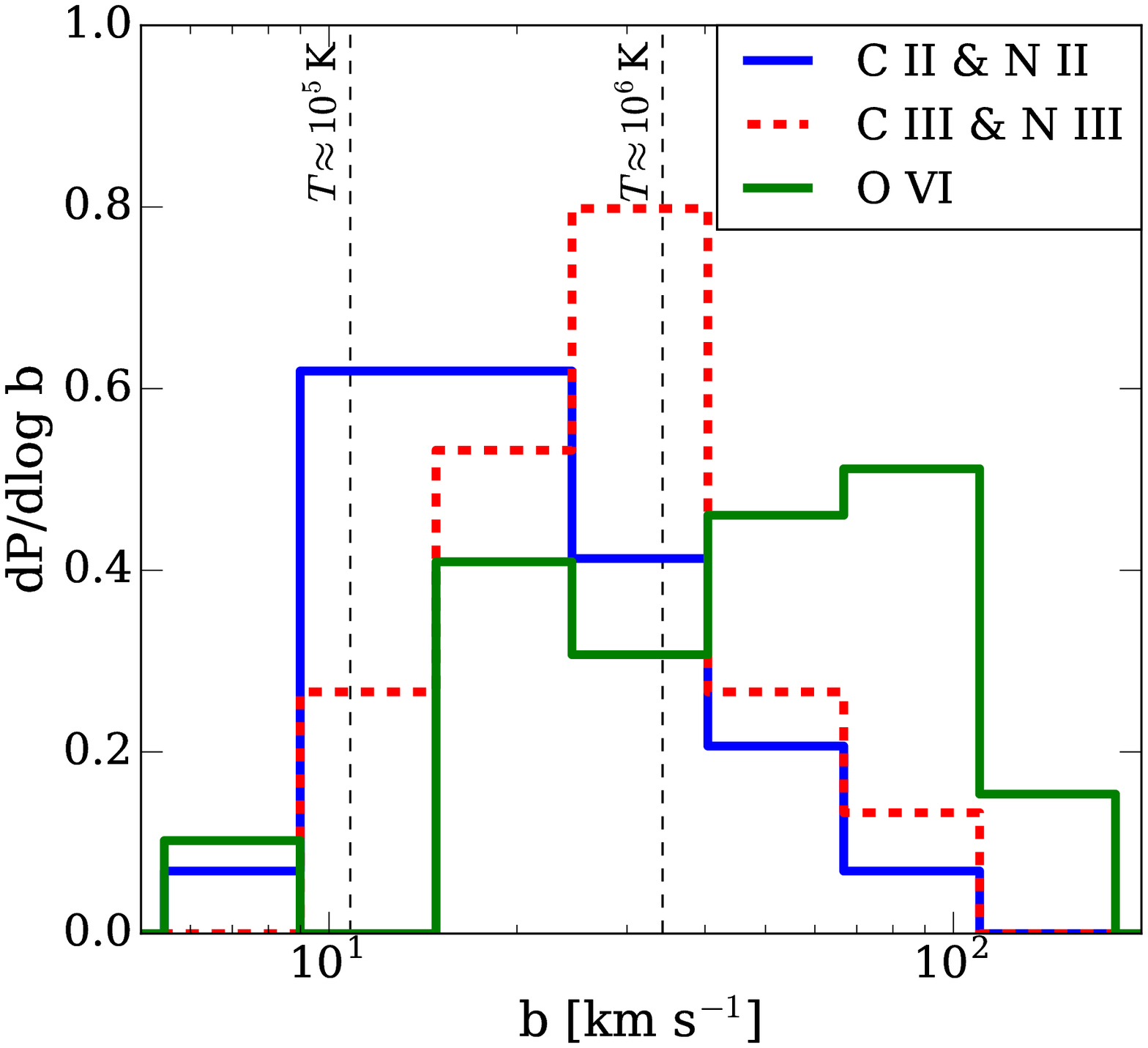, width=5.8cm}
\epsfig{file=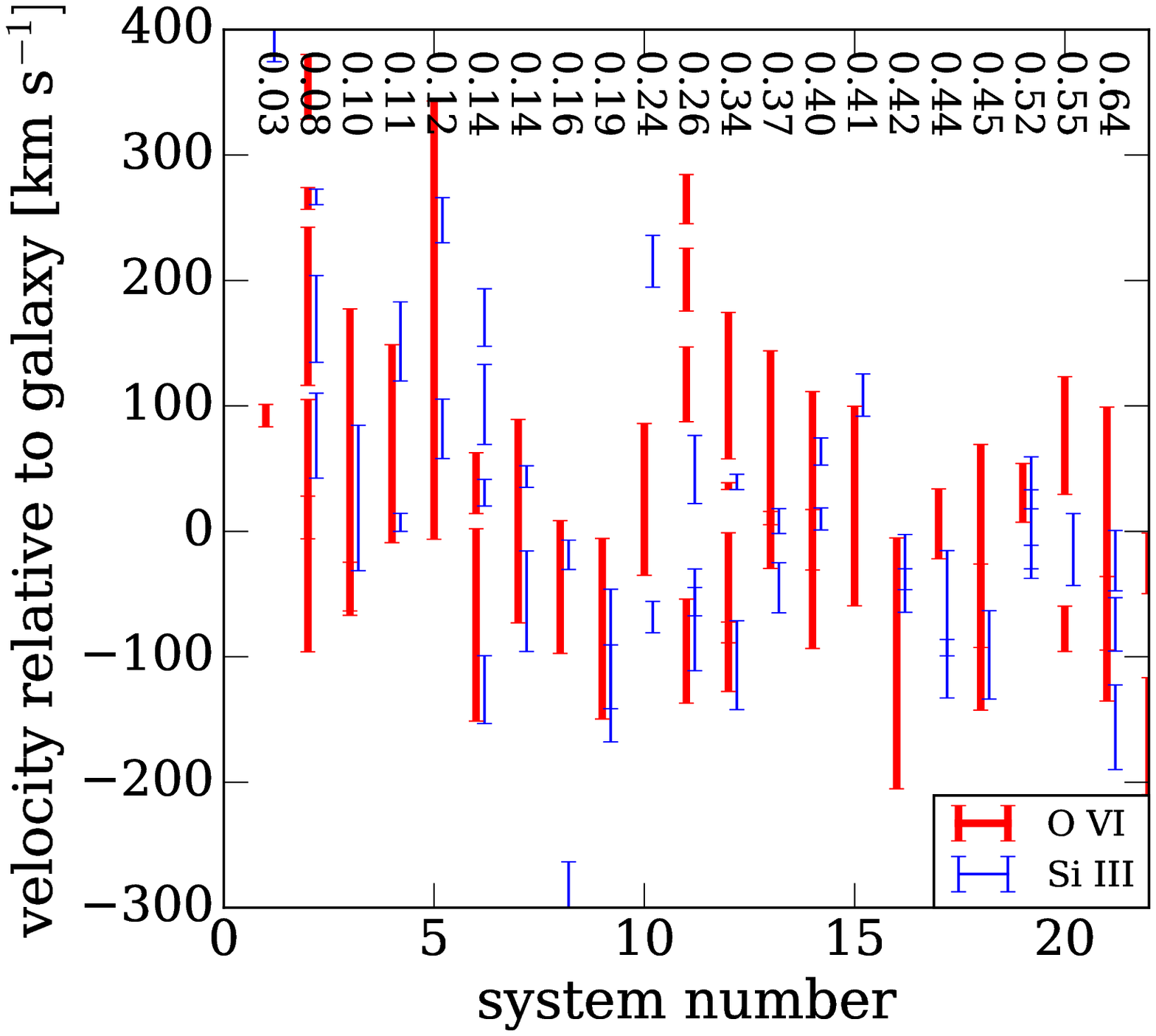, width=5.8cm}
\end{center}
\caption{Summary of pertinent phenomenology for the CGM of $L_*$ galaxies:  All plots use the COS-Halos absorption line measurements \citep{tumlinsonCOS, werk14}.  {\it Left panel:} Solid histograms show the mean column in the specified ionic species at various impact parameters, using the COS-Halos measurements of \citet{werk14}.  To compute a mean in the presence of bounds, all upper limits on the columns were set to zero and all lower limits were set to the $2\sigma$ bound.  Also shown in filled blue circles are the $N_{\rm OVI}$ measurements of `isolated' star-forming galaxies from \citet{johnson15}.  The error bar on each measurement is generally $\lesssim 0.1~$dex.  {\it Middle panel:}  The probability distribution of Doppler b parameters.  The dashed vertical lines are the Doppler widths for $10^5$K and $10^6$K nitrogen gas.  The instrumental resolution for COS of $\approx 18\,$km~s$^{-1}$ has been deconvolved.  {\it Right panel:}  The velocity of \OVI\ (red) and \SiIII\ (blue) for the COS-Halos systems that show absorption in both, sorted by $R/r_{\rm vir}$ (with this value given by the number near the top axis).  The error bars are the lines' b parameters.  \SiIII\ is chosen to represent low ionization metals because it is one of the most ubiquitous metal ions in the $z\sim 0$ CGM.
\label{fig:columns}\label{fig:line widths}\label{fig:velocity_association}}
\end{figure*}

We focus on the HST/COS CGM observations, which are currently our best window into cold and warm gases in the $z\sim0$ CGM. Here we briefly summarize some of the striking features of these observations:

\begin{description}
\item[High \OVI\ columns out to $r_{\rm vir}$:] 
Nearly all sightlines through the halos of star-forming galaxies with impact parameters of $R<r_{\rm vir}$ present projected column densities in \OVI\ of $N_{\rm OVI} = 10^{14.5\pm 0.3}$cm$^{-2}$ \citep{tumlinsonCOS}, with almost no sightlines showing such columns at $R>r_{\rm vir}$ \citep{johnson15}.  The left panel of Figure \ref{fig:columns} plots $N_{\rm OVI}$ as a function of distance in units of the halo virial radius, using the measurements of the COS-Halos project \citep{tumlinsonCOS, werk13} and the stellar mass to halo mass abundance-matching relation of \citet{2010ApJ...710..903M}.  The solid blue histogram shows the mean $N_{\rm OVI}$, and the blue dots show the individual system measurements.  

\item[suprathermal and Gaussian absorption lines:]  The middle panel in Figure~\ref{fig:line widths} shows the probability distribution function of the b parameter for absorption lines of ionic nitrogen, carbon, and oxygen.  The observed line profiles are smooth, with most well fit by one or two Gaussian components.  The dashed vertical lines are roughly the expected b parameter for $10^5$K and $10^6$K gases; most lines have widths well beyond that expected from thermal broadening.  If the \OVI\ traces $10^{5.5}$K gas, the sound speed would be $90\,$km~s$^{-1}$ such that its line widths are largely consistent with turbulent Mach numbers of ${\cal M} \sim 1$.  For the lower ions, however, the inferred Mach numbers would be quite large (${\cal M} = 1-10$) if they trace $10^4$K gas.  It is more likely that these absorbers owe to multiple cloudlets that are entrained in a turbulent hot $\sim10^6\,$K medium.  

\item[coincident line centroids offset from host:]  The velocity-space centroids of the low-ionization lines and of \OVI\ typically coincide with centroid offsets that are smaller than the line widths (see \citealt{werk14, werk16}, and the right panel in Fig.~\ref{fig:velocity_association}).    Only $20\%$ of systems show just \OVI\ and no other metals, and these systems tend to lie at larger projected radii \citep{werk16}.  Additionally, the typical kinematic offsets with respect to the host galaxy are $\Delta v \sim 50-150\,$km~s$^{-1}$, suggesting that most clouds are bound.   Absorption systems along the same sightline are grouped in velocity, rather than multiple lines spanning $\sim 2 \Delta v$ as one would expect for uncorrelated clouds (right panel, Fig.~\ref{fig:velocity_association}).
     
\item[large reservoirs of photoionized gas:]  Additional inferences come from photoionization modeling of the low-ionization metals.  Such modeling infers low densities of at most $10^{-4}-10^{-3}$cm$^{-3}$ and sizes of $\sim 1-10\;$kpc for the $\sim10^4$K photoionized clouds, implying that this gas traces an immense gaseous reservoir of $\gtrsim 10^{10}\Msun$  \citep{stocke13, werk14, stern16}.  Although it has been speculated that this modeling could be skewed by multiphase systems or by additional heating sources, in Section~\ref{sec:HI} we show that for some systems the low inferred densities are unavoidable.
\end{description}

In what follows, we attempt to use the above clues to argue for or against hypotheses for the origin of the \OVI\ gas.

\section{Simple \OVI\ estimates}
\label{sec:simple}
This paper focuses on the large columns of \OVI\ and their association with lower ionization absorbers.  We will argue in ensuing sections that \OVI\ marks gas transitioning through the temperature at which the fraction of collisionally ionized \OVI\ peaks ($T\sim10^{5.5}$K).  If the \OVI\ marks transitioning gas, we can estimate its column density from the timescale over which \OVI\ survives near its peak abundance, $t_{\rm OVI}$, and from the observed radial extent of halo \OVI\ absorption, $R$:
\begin{eqnarray}
N_{\rm OVI} &=& \frac{f_{\rm OVI} [f_{\rm O}]_\odot Z }{\pi R^2 \times \mu_i m_p} \times \dot M \times t_{\rm OVI}, \label{eqn:NOVI_simple1} \\
&=& 2\times10^{14} ~{\rm cm}^{-2}~~R_{200{\rm k}}^{-2} \left(\frac{\dot M}{100 \; \Msun {\rm \, yr}^{-1}}\right) \nonumber \\
&&   \times \left(\frac{Z}{0.3}\right)\left( \frac{t_{\rm OVI}}{\rm 100 \; Myr}  \right),  \label{eqn:NOVI_simple2}
\end{eqnarray}
where  $[f_{\rm O}]_\odot$ is the solar oxygen fraction (set equal to $5\times 10^{-4}$ for eqn.~\ref{eqn:NOVI_simple2} and in ensuing equations; \citealt{2009ARA&A..47..481A}), $f_{\rm OVI}$ is the fraction of oxygen in the \OVI\ ionization state,  $\mu_i\approx 1.2$ is the mean molecular weight of the ions, $\dot M$ is the mass flow rate, $Z$ is the oxygen metallicity relative to solar, and $R_{200{\rm k}}$ is $R$ in units of $200~$kpc. In the spherical collapse model, the virial radius of a $10^{12}\Msun$ halo is $230~$kpc at $z=0.2$, the median redshift of the COS-Halos sample.

There are several timescales that could set $t_{\rm OVI}$.  The first is the isobaric atomic cooling time,
\begin{equation}
t_{\rm cool} = \frac{5 k_b T}{ \Lambda n} \approx 250~ {\rm Myr~} P_{2}^{-1} T_{5.5}^{2.7}  \left(\frac{Z}{0.3} \right)^{-1}, 
\label{eqn:tcool} 
\end{equation}
where $\Lambda$ is defined such that the total cooling rate density is $\Lambda n^2$, and where the second approximate equality is valid for $T=10^5-10^6$K and $P_2\equiv n T/ [100 {\rm \,cm^{-3} K}] > 0.1$ (photoionization increases $t_{\rm cool}$ at lower pressures).  Here, $n$ is the number density of electrons/hydrogen atoms, which to our level of approximation is half the total number density of particles.  
 If $t_{\rm OVI}= t_{\rm cool}$, equation~(\ref{eqn:NOVI_simple2}) shows that generating sufficient \OVI\ columns requires $\dot M/[100 P_{2} \;\Msun {\rm \, yr}^{-1}]\sim 1$. 

{
However, if cooling is too rapid, $t_{\rm OVI}$ could be determined by the timescale over which \OVII\ recombines to become \OVI:
\begin{equation}
t_{\rm rec, OVII\rightarrow OVI} \equiv (\alpha_{\rm OVII\rightarrow OVI} n)^{-1} \approx 130 ~{\rm Myr} ~n_{-4}^{-1} \; T_{5.5},
\end{equation}
where $\alpha_{\rm OVII\rightarrow OVI} $ is the recombination coefficient.  It requires $Z T_{5.5}^{-0.7} > 1$ for $t_{\rm rec, OVI\rightarrow OV} > t_{\rm cool} $ so that $t_{\rm OVI}$ is set by recombinations rather than cooling at solar metallicities and above, somewhat above the metallicities expected in the CGM.  Assuming the oxygen starts in \OVII\ and in the limit of no collisional ionizations, the non-equilibrium \OVI\ fraction is $f_{\rm OVI} \approx \alpha_{\rm OVII\rightarrow OVI}/\alpha_{\rm OVI\rightarrow OV} \approx 0.1$, similar to $f_{\rm OVI} \approx 0.2$ for collisional equilibrium.  Therefore, our conclusions will not change significantly at high metallicities where the \OVI\ is out of equilibrium.  

 }

A final possibility is if the \OVI\ is produced by cooler gas mixing with the virialized $T \sim 10^6$K phase such that the collisional ionization time at the mixed density and temperature, 
\begin{equation}
t_{\rm coll, OVI\rightarrow OVII} \equiv (C_{\rm OVI\rightarrow OVII} n)^{-1} \approx 11 ~{\rm Myr} ~ n_{-5}^{-1} \;T_{6}^{-4},
\label{eqn:tcoll}
\end{equation}
 sets $t_{\rm OVI}$, where $C_{\rm OVI\rightarrow OVII}$ is the collisional ionization coefficient.  This possibility results in much less \OVI\ than if $t_{\rm OVI} \approx t_{\rm cool}$.

 Photoionization could affect our estimates in this section either by slowing cooling or by changing the ionization state of oxygen relative to that in collisional equilibrium.  As mentioned, the cooling time at $T\gtrsim 10^5$K is not affected by the presence of an ionizing background for standard ionizing backgrounds at the pressures we consider (see also Appendix C).  We also find that $f_{\rm OVI}$ is not significantly affected by photoionization at relevant pressures.\footnote{It might be possible that the actual ionizing background (unlike the \citealt{haardt01} model) increased the cooling time more than it decreased $f_{\rm OVI}$. We note that it would not to the extent that \OVI\ is the dominant coolant, and it is one of the primary coolants at $T_{5.5} \sim 1$ \citep{oppenheimer13}.}

Thus, in this picture where \OVI\ is transitioning gas, the observed massive columns of \OVI\ likely result from when $t_{\rm OVI} \approx t_{\rm cool}$ evaluated at $10^{5.5}$K such that equations~(\ref{eqn:NOVI_simple2}) and (\ref{eqn:tcool}) yield
\begin{equation}
N_{\rm OVI} \approx 2.8\times10^{14} {\rm cm}^{-2}~~R_{200{\rm k}}^{-2}P_2^{-1}   \frac{\dot M}{50 \Msun {\rm \, yr}^{-1}}.   \label{eqn:NOVI_tcool}
\end{equation}
To generate the observed $N_{\rm OVI}$ requires fluxes into the \OVI\ phase of tens of solar masses per year, with larger fluxes at higher pressures.  More detailed estimates in the next section show that this simple estimate undershoots the $\dot M$ required to achieve a fixed $N_{\rm OVI}$ by a factor of $\approx 2$.  To maintain approximate steady state, the bulk of this gas is eventually recycled back into the $\sim 10^6$K coronal phase. The galactic feedback power needed to refresh the gas to the virialized phase is
\begin{eqnarray}
\dot E &\approx& \frac{5 k_b T_{\rm vir}}{\mu_i m_p} \dot M, \label{eqn:dotE1}\\ 
&=&  4 \times 10^{49} {\rm erg\;yr^{-1}} \left(\frac{T_{\rm vir}}{10^6{\rm K}}\right) \left( \frac{\dot M}{50 \Msun {\rm \, yr}^{-1}} \right).
\label{eqn:dotE}
\end{eqnarray}
This $\dot E$ estimate is valid for isobaric cooling, and it ignores the power needed to maintain motion in the gravitational potential (which would be roughly equal to $\dot E$ if the cooling gas falls a significant fraction of the virial radius).

{A likely subdominant contribution to the \OVI-sourcing mass flux is recently accreted cooling gas that was shocked at virialization.  Simulations of $z\sim 0$, $10^{12}\Msun$ halos find that they have mean gas accretion rates of $\dot M \approx 8\; \Msun$~yr$^{-1}$ if the gas accretes at the same rate as dark matter \citep{mcbride09} as is likely \citep{2011MNRAS.417.2982F}.  Such $\dot M$ are insufficient to source most of the \OVI\ except at the lowest pressures we argue are possible.}

Cooling gas that has been recycled by feedback to virial temperatures and dispersed on $\sim 100\,$kpc scales likely contributes the bulk of $\dot M$.  Most of the COS-Halos galaxies have star formation rates of $\sim 1-10~\Msun\;$yr$^{-1}$, suggesting supernova rates of $\sim 0.01-0.1~$yr$^{-1}$ with input energies from each supernova of $E_{\rm SN} \sim 10^{51}$erg and, hence, total feedback energies of $\dot E_{\rm kin} \sim 10^{49-50}~$yr$^{-1}$ (and the total energy deposition from massive stars' winds is likely comparable; \citealt{1999ApJS..123....3L}).  Thus, even for the COS-Halos galaxies with the highest star formation rates of $10~\Msun\;$yr$^{-1}$, galactic supernova feedback can support a maximum of $\dot M \sim 100 \; \Msun {\rm \; yr}^{-1}$.  To channel much of the supernova energy into the galactic halo limits cooling losses in the interstellar medium.\footnote{Unclustered supernovae likely lose $\sim90\%$ of their energy by radiative losses in the interstellar medium \citep[e.g.][]{1998ApJ...500...95T, 2014MNRAS.443.3463S}.  Additionally, if cosmic rays or momentum-driven winds power galactic-scale feedback, these mechanisms' energetics budgets are capped at $\sim10\%$ the total energetics from supernovae \citep[e.g.][]{2005ApJ...618..569M, 2008ApJ...687..202S} and so are disfavored by the high $\dot E$ required if the \OVI\ is cooling gas.}

Energetics considerations bound the thermal pressure in the \OVI\ region to $P \lesssim 100\,$cm$^{-3}$K as saturating this bound requires the total available energy in supernova feedback.  At the same time, cooling within the age of the Universe from the virialized phase with temperature $T_6$ -- as is necessary to generate the \OVI\ in the above scenario -- requires $$P> 30 T_6^{-2.7} (Z/0.3)^{-1} \,{\rm ~cm^{-3}~K}.$$  Lower pressures are also disfavored by total mass arguments:  In the spherical collapse model, the density of gas within a halo's virial radius is $n = 1.0\times10^{-4}\,$cm$^{-3}$ at $z=0.2$, or a pressure of $P=100\,$cm$^{-3}$K for $T=10^6$K, assuming the closure fraction in baryons.  Thus, $P=10\,$cm$^{-3}$K at all radii would result in only a tenth of the baryons -- roughly $10^{10}\Msun$ for a $10^{12}\Msun$ halo -- within the virial radius, which equation~(\ref{eqn:intro}) suggests is the absolute minimum gas mass required to produce the \OVI\ for solar abundances.  In more physically motivated scenarios, we find that more gas than $10^{10}\Msun$ is required (Section~\ref{ss:coolingflow}).  
  
Thus, these arguments suggest that on average the halo gas has pressures of  $$10 {\rm \; cm^{-3}\,K}< P \equiv n T < 100 {\rm \; cm^{-3}\,K}.$$ These bounds on the pressure are in accordance with constraints from the Milky Way (as discussed in the introduction) and with densities found in CGM simulations \citep{feldmann13, fielding16}.\footnote{While the relations in this section essentially do not rely on the evolution of cooling gas being isobaric or isochoric, our discussion of the allowed pressures suggests the isobaric case.  In the absence of nonthermal pressure, isobaric conditions are satisfied on scales shortward of the distance a sound wave travels in a cooling time, or
$t_{\rm cool} c_s =  20 ~{\rm kpc~}  P_{2}^{-1} T_{5.5}^{3.2} \times [0.3/Z]$.
}

While our estimates for $N_{\rm OVI}$ and $\dot E$, equations~(\ref{eqn:NOVI_tcool}) and (\ref{eqn:dotE}) do not depend on the gas metallicity, $Z$, others do.  Here we comment briefly on what metallicities are reasonable. \citet{peeples14} estimated that a galaxy produces $(0.03-0.1) M_\star$, where $M_\star$ is the galaxy stellar mass, which translates to $Z = 0.1-0.4$ if all the metals are in the CGM, if the CGM contains all the halo's baryons, and if $M_\star/[M_{\rm halo} \Omega_b/\Omega_m] = 0.1$ in accord with abundance-matching estimates for $L_*$ galaxies \citep{2010ApJ...717..379B}.  Of course, we do not expect all gas to be well-mixed, and indeed the photoionized absorbers discussed in Section~\ref{sec:HI} may have higher metallicities than average. \citet{prochaska17} measured the metallicity of these photoionized CGM clouds and found $Z = 0.02-3$ at $95\%~$C.L., with a median of $Z = 0.3$. We adopt $Z=0.3$ as our fiducial metallicity throughout.

\section{Hypotheses for process that creates CGM \OVI}
\label{sec:hypotheses}
This section considers specific hypotheses for what generates the CGM \OVI\ and discusses how these hypotheses are either consistent or not with the observed CGM.  

\subsection{the \OVI\ reflects flows of cooling gas}
\label{ss:coolingflow}

\begin{figure*}
\begin{center}
\epsfig{file=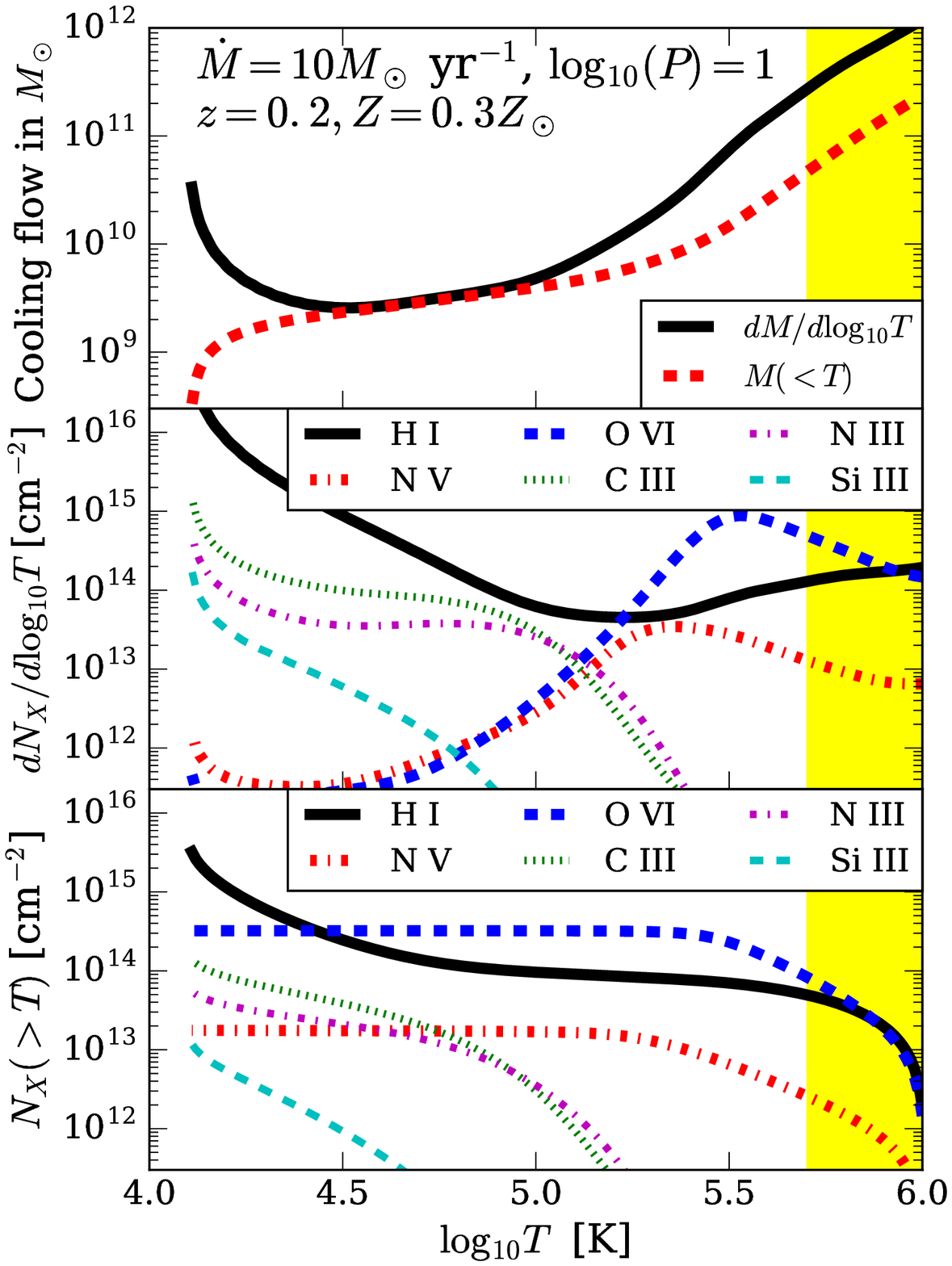, width=8cm}
\epsfig{file=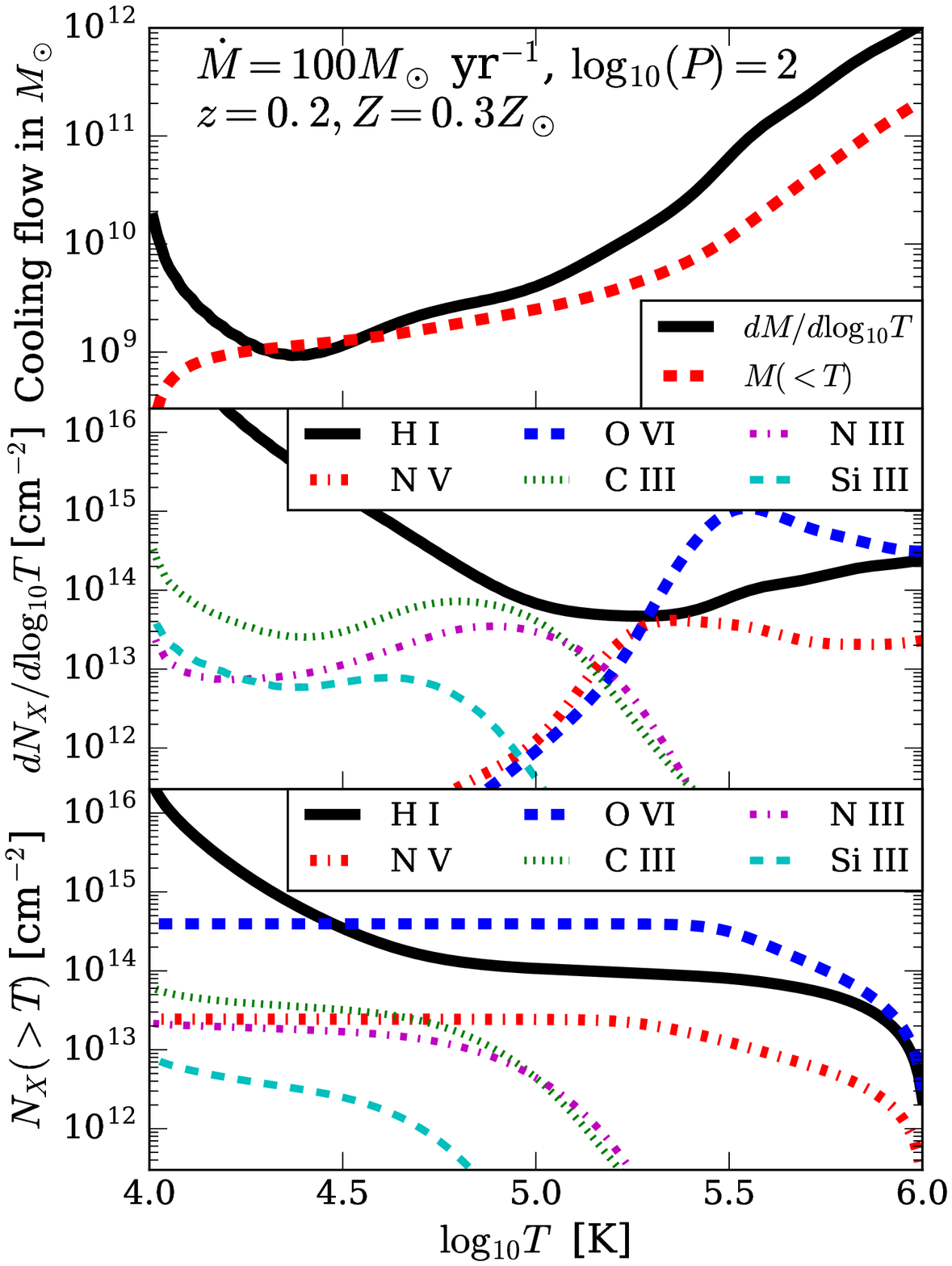, width=8cm}
\end{center}
\caption{Cooling gas calculations where the parameters are chosen to yield $N_{\rm OVI} \approx 10^{14.5}$cm$^{-2}$, with specifications given by the text in the top panels. The gas/column for all calculations is assumed to have a top-hat angular profile with $R=200\,$kpc, and non-equilibrium chemistry is followed assuming isobaric cooling and the \citet{haardt01} ionizing background model.  The {\it top panels} show the fraction of mass per $\log_{10} T$ as well as the cumulative mass up to a given temperature that is cooling, assuming isobaric conditions. The {\it middle panels} show the column density per $\log_{10} T$ in select ions for the same specifications as given in the corresponding top panel.  Note that $N_X = \int d\log_{10} T ~d N_X/d\log_{10} T$, such that the linear area under the curves gives the column. The {\it bottom panels} show the cumulative column that resides in gas above the specified temperature.  {The yellow shaded region in the left panel delineates the maximum temperature that yields a consistent calculation:  The shaded higher temperature gas would not fit within $r_{\rm vir}$ at $P=10~$K~cm$^{-3}$ (see the main text).}
\label{fig:TI}}
\end{figure*}

This paper's thesis is that the \OVI\ owes to cooling gas driven by thermally instability in the halo atmosphere.  The previous section calculated how much of a flow is required to generate the observed \OVI\ columns; equations~(\ref{eqn:NOVI_simple2}) and (\ref{eqn:tcool}) suggest $\dot M \sim 100 \,\Msun$~yr$^{-1}$ for $P_2 = 1$ is needed, with the scalings
\begin{equation}
N_{\rm OVI} \propto \dot M P^{-1} {\rm ~~and~~} M  \propto  \dot M P^{-1} Z^{-1} \propto N_{\rm OVI} Z^{-1}.\nonumber
\end{equation}
Thus, lower accretion rates ($\dot M$) require smaller pressures ($P$), and, fixing $N_{\rm OVI}$, lower total mass participating in cooling ($M$) requires higher metallicities.  Figure~\ref{fig:TI} shows a more detailed calculation of the fraction of mass per $\log_{10} T$ (top panels) and of the average column per $\log_{10} T$ (middle panels) for isobaric cooling with $P \equiv n \,T = 10~{\rm K~cm}^{-3}$ (left panels) and $P = 100~{\rm K~cm}^{-3}$ (right panels).  The mass flow rate has been tuned in the respective panels to $\dot M = 10\Msun {\rm ~yr}^{-1}$ and $\dot M =100\Msun {\rm ~yr}^{-1}$ in order to yield $N_{\rm OVI} \approx 3\times 10^{14}~$cm$^{-2}$ in accord with observations.  These calculations assume a cylindrical gas profile with angular extent of $R =200~$kpc, with the line-of-sight projected extent adjusted to accommodate the specified parameters.    They require a factor of $\approx 2$ larger $\dot M$ to achieve a specific $N_{\rm OVI}$ compared to the simpler estimates in the last section.
   
 The top panels in Figure~\ref{fig:TI} show the cumulative mass below $T$ in the cooling material (dashed curve) as well as its logarithmic derivative (solid curve).  The mass that cools from $\sim 10^6$K must be $M \sim10^{11}~\Msun$, comparable to the baryonic mass associated with these halos.    If purely isochoric conditions exist between $10^{5.5}$K and $10^6$K, which is less likely than at lower temperatures, the required $M$ a fixed $N_{\rm OVI}$ and $\dot M$ would be a factor of $\approx 10^6/10^{5.5}\times 3/5 \approx 2$ smaller if the gas starts out at the same density. 
 
{  An apparent inconsistency with the calculation in Figure~\ref{fig:TI} is that at the virial temperature and $P=10~$K~cm$^{-3}$, only $M \approx 2\times10^{10}~\Msun$ can fit within the virial radius (Section~\ref{sec:simple}), much smaller than the $M \sim10^{11}~\Msun$ the cooling gas flows model requires for $\sim 10^6$K.  The shaded yellow regions in Figure~\ref{fig:TI} indicate the temperatures above which the amount of mass exceeds that which can be contained.  For  $P=10~$K~cm$^{-3}$, this temperature corresponds to $10^{5.7}$K and, for $P=100~$K~cm$^{-3}$, to $10^{6}$K.  There are two ways out of this inconsistency:  (1) the gas must start cooling from temperatures below these values, or (2) the hotter gas (which we do not have solid observational probes for) is distributed on a larger scale and then migrates inward by the time it is observed in \OVI.}


In this cooling material picture, there are three possibilities for what drives the kinematics: (1) condensation owing to loss of buoyancy, (2) thermal instability occurring in outflowing gas, or (3) large-scale sloshing motions (i.e. coherent bulk motions).\footnote{Large-scale sloshing motions are also seen in X-ray maps of the cores of galaxy clusters \citep{2014ApJ...781....9G}.}  We favor the latter since in the other possibilities it is less clear why there should be an order one covering fraction and why clouds should be so kinematically correlated.  However, while most relative velocities with respect to the galaxy are $<100~$km~s$^{-1}$, some reach $\sim 300~$km~s$^{-1}$, which are too large to reasonably owe to sloshing.  Instead, we suspect they owe to outflows (which is supported by some of the highest kinematic offsets being at the smallest $R$; Fig.~\ref{fig:velocity_association}).  Appendix~A discusses various kinematic timescales.  

In the sloshing picture, the cold and warm absorbers are comoving in the hot halo atmosphere, with the observed velocity offsets from the host galaxy due to sloshing of the entire atmosphere.  The energy dissipation from turbulence driven by these sloshing motions is
 \begin{eqnarray}
 \dot E_{\rm slosh} &\sim& \frac{1}{2} M v^2 t_{\rm sc}^{-1}, \label{eqn:sloshing}\\
  &=& 1\times10^{49}~ {\rm erg~ yr}^{-1}~  v_2^2 R_{200k} \left( \frac{M}{10^{11} \Msun}\right) \left( \frac{T_{\rm vir}}{10^{6} {\rm K}}\right)^{-1/2}, \nonumber
 \end{eqnarray}
 where $v$ is the characteristic velocity of sloshing material and $t_{\rm sc}$ is the halo's sound-crossing time for virialized gas (which is also the dynamical time).  $ \dot E_{\rm slosh}$ is comparable to the energy needed to reheat the cooling gas (e.g. eqn.~\ref{eqn:dotE}).  Co-motion with the sloshing atmosphere for the photoionized clouds  requires either cloud lifetimes that are much shorter than the $\sim 1~$Gyr free-fall time from $\sim r_{\rm vir}$ and/or for the clouds to have similar densities to the virialized gas such that they are largely entrained.  We find in Section~\ref{sec:HI} that the cloud lifetimes could be much shorter than the free-fall time.  We also find in that section that the densities of the \HI\ clouds appear to be just $\sim 10\times$ that of the virialized gas.  In this case, the cloud terminal velocity can be less than $100~$km~s$^{-1}$  (Appendix A) and, hence, they may appear as being largely entrained.

The cooling material picture predicts that other ions than \OVI\ should be observable, especially if the cooling gas is not disrupted before it reaches lower temperatures.  The bottom panels in Figure~\ref{fig:TI} plot the cumulative column that resides at temperatures above $T$ for several ions. \NV\ traces very similar temperatures to \OVI\ and should appear with $N_{\rm NV}/N_{\rm OVI} \approx 0.1$, consistent with current upper bounds from COS-Halos, which fall just above this ratio for most systems.  (\NV\ has been detected in three COS-Halos absorption systems to be \NV/\OVI\ $ = 0.05$, $0.16$, and $0.15$; \citealt{werk16}.)   Lower ionization metals, such as \NIII\ and \CIII, show columns of $\sim 3\times10^{13} $cm$^{-2}$ from $T>10^{4.5}$K gas in the cooling flow.  [Note that there is a buildup of gas that diverges in the limit of the equilibrium temperature at $\sim 10^4$K:  the actual amount of gas at these temperatures depends on the survival time of photoionized clouds (Section~\ref{sec:HI}).]  The columns in \NIII, \CIII, and \SiIII\ are $\sim 10\times$ smaller than the mean observed column at  $T>10^{4.5}$K gas and so the observed columns in these ions are unlikely to be from relatively high-temperature cooling gas.  Lastly, gas in the flow of cooling material produces \HI\ with $N_{\rm HI} \sim 10^{14-15}$cm$^{-2}$ for $T\gtrsim10^{4.5}$K.  Such columns may be able to explain the \HI\ columns of $N_{\rm HI} \approx 10^{14.5}$cm$^{-2}$ and $N_{\rm HI} \approx 10^{15.1}$cm$^{-2}$ for the two COS-Halos systems in which there is \OVI\ absorption but none identified from other metal ions \citep{werk14}.\footnote{Circumgalactic dust provides another constraint on the cooling flows picture.  Dust appears to survive in mass in the CGM \citep{menard10, 2015ApJ...813....7P}, even though in the cooling flows model sputtering may destroy it.  The destruction timescale in $10^6$K gas is is $\tau_{\rm dist} \approx  20 {\rm ~~Gyr~~}  \left(\frac{a_d}{0.05 \rm \mu m} \right) n_{-5}^{-1}$, 
where $a_d$ is the size of the grain \citep{1979ApJ...231...77D}, with the observations suggesting $a_d\approx 0.05 \;\mu m$ grains \citep{2015ApJ...813....7P}.  This timescale is a strongly increasing function of temperature at $T<10^6$K and has little dependence at higher temperatures.  It also depends on the uncertain material properties of grains.  At the lowest densities we argue for of $n_{-5}=1$, $\tau_{\rm dist}$ is longer than the age of the Universe and so most dust should survive.  For the highest densities we argue for, $n_{-5}=10$, $\tau_{\rm dist}$ would suggest significant destruction and may not be consistent with the substantial observed dust columns.}

\paragraph{Conclusion:}  Flows of cooling gas could explain the $N_{\rm OVI}$ if $M \sim 10^{11}\Msun$ is participating in the flow with $\dot M \sim 100 P_2 \Msun~$yr$^{-1}$.  That much of the halo gas is cooling in this scenario provides an explanation for the large covering factors and the velocity coherence of CGM absorbers.  This picture requires quiescent galaxies to have lower $\dot M$ to explain their reduced incidence of \OVI.

\subsection{the \OVI\ owes to boundary layers, shocks and hot winds}
\label{ss:other}
This section briefly discusses why the bulk of the \OVI\ \emph{cannot} owe to gas in boundary layers, shocks in $10^4$K clouds, and hot winds.  
  In each of these scenarios, there is a flow of gas with flux $v \,n$ that proceeds to cool with timescale $t_{\rm cool}$, yielding a column of \citep[e.g.][]{heckman02}
\begin{eqnarray}
N_{\rm OVI}^{\rm system}  &= &  f_{\rm OVI}\, \eta \, [f_{\rm O}]_\odot \, Z\, v \, n \,t_{\rm cool},\\
    &\approx& 7\times 10^{14} {\rm ~cm}^{-2} ~ \eta \; v_{2}, 
    \label{eqn:NOVIss}
\end{eqnarray}
where $v_2 \equiv v/[100 \;{\rm km~s}^{-1}]$, $\eta$ is an order unity fudge factor, and equation~(\ref{eqn:NOVIss}) uses equation~(\ref{eqn:tcool}) for $t_{\rm cool}$ evaluated at $T_{5.5} = 1$ where $f_{\rm OVI} \approx 0.2$.  This rough estimate applies in multiple situations:  to stable turbulent boundary layers (where, in addition to geometric factors, $\eta$ can be thought of as a mixing efficiency), to shocks (where $\eta^{-1}$ is the shock compression in the \OVI-bearing region), or to winds (where $\eta$ is a suppression factor due to adiabatic losses).  Nicely, $N_{\rm OVI}^{\rm system}$ is insensitive to the overall density and metallicity and is comparable to the observed \OVI\ columns for reasonable velocities.\footnote{Equation~(\ref{eqn:NOVIss}) predicts a relationship between the \OVI\ column and velocity \citep{heckman02}.  If we take $v$ to be the b-parameter of the line, this relationship appears to be present in the COS-Halos data, especially for systems that show coincident low ionization absorption \citep{werk16}, and also for Galactic and even intergalactic \OVI\ systems \citep{heckman02, 2016arXiv160507187B}.  Rather than indicating that these mechanisms are at play in the CGM, we think the observed relationship is a coincidence and instead reflects the tendency for higher column systems to exhibit broader lines.}   

The picture that yields equation~(\ref{eqn:NOVIss}) also predicts a width for the \OVI\ region:
\begin{eqnarray}
L^{\rm system}_{\rm OVI} \sim v \,t_{\rm cool} \approx 30 {\rm ~kpc~} v_{2} \left({0.3}/{Z}\right) P_{2}^{-1}.
                   \label{eqn:Lss}
\end{eqnarray}
Thus, the widths for the \OVI--bearing gas in this picture are quite broad, especially when accounting for our energetics bound $P_{2} <1$ (Section~\ref{sec:simple}).

With these general estimates in mind, let us now discuss the specific cases of turbulent boundary layers, shocks, and hot winds.  While all of these possibilities could explain a component of the \OVI\ absorption, they fail generally in that they cannot explain its vast scope.

\paragraph{turbulent boundary layers:}  Turbulent boundary layers should naturally form around moving cold clouds as eddies mix their gas with the hot atmosphere, generating a mixed phase with temperature near the geometric mean of the two for isobaric conditions \citep{begelman90}.  This picture would provide an explanation for the kinematic association of the warm and colder gases.  Numerical calculations show that $\sim 10$ boundary layers would be required to explain the observed $N_{\rm OVI}$ \citep[suggesting $\eta\sim 0.1$ in eqn.~\ref{eqn:NOVIss};][]{kwak10}.   For $P_2\lesssim 1$, the boundary layer is broad with a characteristic width of $L_{\rm BL} \sim 30 P_{2}^{-1}$kpc (eqn.~\ref{eqn:Lss}).  Thus, turbulent boundaries would have to be of comparable size to, or even larger than, the $\sim 1\,$kpc clouds that are being mixed. (Section~\ref{sec:HI} estimates the line-of-sight cloud sizes.)  
  Generating the bulk of the \OVI\ with turbulent mixing requires broad cooling regions that include $\gtrsim 10^{10} \Msun$ and that fill much of the halo such that the picture would be of large-scale mixing rather than boundary layers.  The similarity of the global mixing picture with the flows of cooling gas model means we henceforth do not distinguish between the two scenarios.\footnote{Since the mixing owes to shear, the simplest boundary layer model predicts that there is a correlation between the b parameters of the lines (which vary quite substantially) and the projected velocity of the cloud relative to the host galaxy.  We find no evidence for such a correlation in the COS-Halos data.}

\paragraph{shocks in $\sim 10^4$K clouds:} Cooling gas behind fast shocks within the $10^4$K photoionized clouds would generate warm \OVI-bearing gas.  Such shocks may be driven by, for example, galactic outflows buffeting the cold clouds.\footnote{To generate collisionally ionized \OVI, the shock must heat gas to $T^{\rm shock} \gtrsim 10^{5.5}$K, which requires shock velocities of $v_s =150 {\rm ~km~s^{-1}} (T^{\rm shock}_{5.5})^{1/2}$ and the cold clouds must be buffeted by flows in the hotter intrahalo gas with flow velocities that are even larger by the square root of the density difference between the hot gas and photoionized clouds. These shock velocities are faster than the shocks created in the process of restoring pressure balance.
\label{vhot}}   Assuming the gas is able to cool behind the shock, the extent of the \OVI-bearing downstream region for a shock with velocity $v$ is $L_{\rm shock} \sim 30 v_{2} (Z/0.3) P_{2}^{-1}~$kpc, where $P\equiv nT$ gives the pressure of the shocked gas -- enhanced relative to the ambient gas by ${\cal M}^2 \times [10^4 {\rm K}/10^{5.5} {\rm K}] \gtrsim 3$ with ${\cal M}= 10$ for shocking to $T_{5.5}=1$ and a purely hydrodynamic shock.  We remind the reader of our energetics bound of $P_2<1$ (eqn.~\ref{eqn:dotE}).  Such $L_{\rm shock}$ are larger than the photoionized clouds the shocks are penetrating (Section~\ref{sec:HI}) so that each cloud is likely destroyed by the shock and would not generate the full column estimated by equation~(\ref{eqn:NOVIss}).  Such shocking would again have to heat at least $10^{10}\Msun$ of gas to \OVI\ bearing temperatures, comparable to our inferences for the total amount of $\approx 10^4$K gas that is available to shock (Section~\ref{sec:HI}).  Furthermore, multiple shocks that are moving in various directions would be required to generate the broad Gaussian lines centered on the lower ionization absorbers \citep{werk16}.  This multiple-shock scenario would generate line widths that are too broad, $\gtrsim300~$km~s$^{-1}$, if the shocks are sufficiently fast to heat the $10^4$K gas to \OVI-bearing temperatures of $T_{\rm shock} \gtrsim 10^{5.5}$K (see footnote~\ref{vhot}).  We conclude that shocks penetrating $\sim 10^4$K gas cannot explain the bulk of the \OVI.

\paragraph{hot winds}
\OVI\ could arise from winds sourced by galactic star formation.  If these winds start out kinematically cold, then the \OVI\ they produce must owe to the material later mixing and shocking as considered previously.  If they instead start hot with $T > 10^{5.5}$K, and cool through this temperature, they would more directly produce \OVI.  If the cooling is dominated by atomic processes rather than by adiabatic cooling, the \OVI\ column is given by equation~(\ref{eqn:NOVIss}) with $\eta \sim 1$, and the wind requires $\dot M \sim 10 P_1 \Msun {\rm \, yr}^{-1}$ to explain the observed \OVI\ columns (eqn.~\ref{eqn:NOVI_tcool}) with $1 < P_1< 10$.  [\citet{thompson16} found that if the mass loading rates are $\dot M/SFR \gtrsim 0.5$, the wind will eventually be dominated by atomic cooling. However, the $\dot M/{\rm SFR}$ in the \citealt{1985Natur.317...44C} hot wind model is limited to be $\dot M/{\rm SFR}\lesssim 1$ to not violate constraints on X-ray emissions \citep{zhang14, meskin16}, in strong tension with the $\dot M \sim 10 P_1 \Msun~$yr$^{-1}$ we require.]  Leaving aside the fact that such massive outflow rates cannot be sourced by $L_*$ galaxies, another major difficulty with the \OVI\ being created in a hot wind owes to the large spatial extent of observed \OVI:  models find that the wind should cool through $10^{5.5}$K over a relatively narrow range in radius.  For example, \citet{thompson16} added atomic cooling to the adiabatic wind model of \citet{1985Natur.317...44C}.  They found that this wind model's \OVI\ absorption occurs at $r\lesssim 15\,$kpc.  Thus, hot winds cannot explain the extended \OVI\ halos.  

\subsection{the \OVI\ is photoionized and in thermal equilibrium}
\label{sec:photoionized}  

There are two significant appeals of making the \OVI\ photoionized gas in thermal equilibrium, with the heating dominated by photoionization.\footnote{In Section~\ref{sec:HI} we show that large heating rates above photoionization would be required to raise the temperature well above the temperature in photoionization equilibrium.}  The first is that this scenario avoids the large flux of gas through unstable temperatures.  The second is that less tuning is required for the oxygen to be in the \OVI\ ionization state compared to if it were collisionally ionized.  Namely, the \OVI\ abundance has $f_{\rm OVI} \approx 20\%$ for over the broad range of $n=3\times 10^{-6} - 3\times10^{-5}\cc$ (baryonic overdensities of $10-100$ at $z=0.2$) for standard ultraviolet background models  and $T\lesssim 10^5$K \citep{stern16}.  This broad range of densities and temperatures is in contrast to collisional ionization equilibrium, where $f_{\rm OVI}$ peaks at $\approx 20\%$ only for a narrow peak in temperature centered at $T= 10^{5.5}$K.

However, the difficulties with \OVI\ being photoionized and in thermal equilibrium are immense.  First, to explain the \OVI\ absorbers with such low-density $T\sim 10^{4.5}$K gas\footnote{$T\sim 10^{4.5}$K  is roughly the equilibrium temperature for $n=10^{-5}{\rm cm}^{-3}$ and $Z=0.3$.} requires it to constitute a significant volume of $L_*$ galaxy halos with spatial extents of
\begin{eqnarray} Òthe flows of the cooling gasÓ
 L_{\rm OVI} &\equiv& \frac{N_{\rm OVI}}{f_{\rm OVI} [f_{\rm O}]_\odot Z n}, \\
 &=&  100 {\rm\, kpc~} Z^{-1} n_{-5}^{-1}  \left(\frac{N_{\rm OVI}}{10^{14.5} {\rm cm}^{-2}} \right) \left(\frac{0.2}{f_{\rm OVI}} \right),
 \label{eqn:sizeOVI}
\end{eqnarray}
again noting that $0.3 \lesssim n_{-5} \lesssim 3$ for $f_{\rm OVI}\approx 0.2$.  (Furthermore, observations of the ratio \NV/\OVI\ require $n \lesssim 10^{-5}$cm$^{-3}$ for a solar [N/O] abundance ratio; \citealt{werk16}.)   Second, the extremely low densities of photoionized \OVI\ absorbers in turn require low thermal pressures of $P=0.1-1~$K~cm$^{-3}$.  The low thermal pressures imply either an extremely small mass in the canonical $\sim 10^6~$K virialized component (resulting in such low densities that it cannot cool in the age of the Universe) or that the \OVI-bearing clouds are primarily supported by nonthermal pressure (from magnetic fields or cosmic rays).   For the former scenario, the \OVI\ absorbers cannot collide and re-form a  pressurizing virialized atmosphere, as would be naively expected on a short timescale of $(10^{4.5} {\rm K}/T_{\rm vir} /f_{\rm cov}) \times t_{\rm dyn}$ in the absence of coherent motion, where $f_{\rm cov}\sim 1.5$ is the ``cloud'' covering fraction \citep{stocke13} and $t_{\rm dyn}$ is the $\sim 1$ Gyr halo dynamical timescale.  For both scenarios, the required gas densities of the \OVI\ gas are well below those needed to generate significant columns in the lower ions, implying that this gas coexists with gas at other densities \citep{stern16}.\footnote{A large enhancement in the ionizing background at the ionization potential for \OVI\ ($138\;$eV) over standard ionizing background models would alleviate the need for small $P$ \citep{werk16}.  A factor of $\gtrsim 100$ increase would be required to be consistent with Galactic observations that suggest $P= 100\,$K~cm$^{-3}$.  In Appendix~\ref{sec:proximity} we argue that more than a factor of few enhancement is unlikely.}

\paragraph{Conclusion}  Photoionization of \OVI\ requires either a vastly underpressurized CGM or one in which nonthermal pressure dominates over the thermal pressure of \OVI-tracing gas by a factor of $\sim 100$.   This unlikely scenario also requires an unlikely ${\cal O}(1)$ fraction of the volume of $L_*$-galaxy halos to be filled with $\sim 10^{4.5}~$K gas.

\subsection{the OVI is virial temperature gas}
\label{ss:Tvir}

The final possibility we consider is for the \OVI\ to owe to halo gas near the halo virial temperature of $T_{\rm vir} \sim 10^6$K  \citep{oppenheimer16}.  If the \OVI\ owes to virialized gas, there is a natural explanation for the deficit of \OVI\ in quiescent galaxies -- quiescent galaxies likely have higher virial temperatures and, hence, less \OVI\ \citep{oppenheimer16}.  This scenario also seems consistent with the comparable columns of \OVI\ at all impact parameters out to a virial radius -- the natural extent of virialized gas (e.g.~Fig.~\ref{fig:columns}).

\begin{figure}
\begin{center}
\epsfig{file=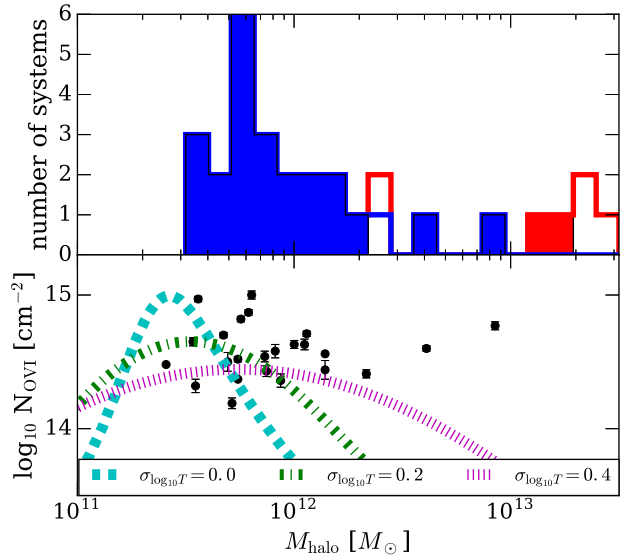, width=9cm}
\end{center}
\caption{{\it Top panel: }Histograms of the halo mass for the COS-halos sample of \OVI\ absorbers.  The halo mass is estimated from abundance matching to convert the observed $M_\star$ to $M_{\rm halo}$.   The filled histograms include only systems with \OVI\ detections, and the unfilled histograms include all observed absorption systems.  The blue histograms are restricted to galaxies with specific star formation rates of sSFR $> 10^{-11}$yr$^{-1}$, whereas the red histograms require sSFR $< 10^{-11}$yr$^{-1}$.  {\it Bottom panel:} The points with error bars show the \OVI\ column for all the detections for star-forming galaxies (sSFR $> 10^{-11}$yr$^{-1}$).  The cyan dashed line shows the expected column if all of the gas has temperatures of $T_{\rm vir}$, assuming the gas is a top-hat out to the virial radius, and the two other curves are the same except that they take a lognormal distribution of temperatures peaked at $T_{\rm vir}$ with the quoted standard deviation.
 \label{fig:mhaloTvirCOS}}
\end{figure}

However, the amount of \OVI\ is extremely temperature sensitive, with $ f_{\rm OVI} = 0.02, 0.003, 0.001$ for  $T_{\rm vir} = \{0.5, 1, 2 \} \times 10^6{\rm \, K}$ for $n=10^{-4}~{\rm cm}^{-3}$ and the \citet{haardt01} ionizing background.  Yet, the virial temperature of halos ($\propto M_{\rm halo}^{2/3}$) likely varies substantially across the COS-halos sample.  How can $N_{\rm OVI}$ be so uniform across this sample?  The top panel in Figure~\ref{fig:mhaloTvirCOS} shows histograms of halo mass from the COS-halos sample, estimated from abundance matching to convert $M_\star$ to $M_{\rm halo}$. The filled histograms include only systems with detected \OVI, and the unfilled histograms include all absorption systems.  The blue histograms show star-forming galaxies, and the red ones show quiescent galaxies, with the division set at a specific star formation rate (sSFR) equal to $10^{-11}$yr$^{-1}$.  In the bottom panel, the points with error bars show the COS-Halos measurements of $N_{\rm OVI}$ for all the star-forming galaxies.   The dashed cyan curve shows the predicted $N_{\rm OVI}$ if all the halo gas has a temperature of $T_{\rm vir}(M_{\rm halo})$, using the form predicted in spherical collapse plus an isothermal potential out to the virial radius (e.g. \citealt{barkana01}).\footnote{An isothermal potential out to the virial radius underestimates the characteristic temperature at most radii, and so, for a more realistic potential, the $T=T_{\rm vir}$ model would be in even more tension with the data.}    This curve is also calculated for $n=10^{-4}{\rm cm}^{-3}$ -- the \OVI\ column would decrease for lower $n$ as photoionization becomes more important.  This curve fails to match the lack of correlation between $N_{\rm OVI}$ and $M_{\rm halo}$ observed in star-forming galaxies.

To reduce the strong predicted trends in $N_{\rm OVI}(M_{\rm halo})$ requires each virialized corona to have gases with a distribution of temperatures.  Some temperature spread is expected as there is no reason for all of the gas temperatures to be exactly $T_{\rm vir}$.  The other two curves in the bottom panel of Figure~\ref{fig:mhaloTvirCOS} model this spread as a lognormal distribution peaked at $T_{\rm vir}$ and with the quoted standard deviation (see \citealt{faerman16} for a related model).  To maintain similar columns across the observed range of $M_{\rm halo}$ for star-forming galaxies, the distribution of temperatures would have to be comparable to the factor of few spread in halo masses in the COS-Halos sample.  These lognormal models still underpredict the typical column by a factor of two, suggesting more gas is required at $T\sim 10^{5.5}{\rm K}$.  The models fail by a factor of $>5$ to explain the \OVI\ columns in the two star-forming systems with the highest inferred halo masses.  

Physically, a spread in temperature would likely be achieved dynamically, as the lower the temperature, the quicker gas cools ($t_{\rm cool} \sim T^{-2.7}$) such that heating is less likely to balance cooling with decreasing temperature.  A potential scenario for generating a range of temperatures without most of the cooling gas running away to temperatures of $T\sim 10^4$K is for the cooling gas to be reheated by shocking and other feedback processes to a relatively high temperature.  The $\dot M$ that is required in this scenario should be similar to that in the cooling flows model, in which we inferred $\dot M =100 P_2 \Msun {\rm yr}^{-1}$.  The issue that we leave unresolved is whether it is more physical for the cooling flows to be disrupted at \OVI-bearing temperatures or at the lower temperatures characteristic of thermal equilibrium.  Both phases may have similar lifetimes:  We infer survival times for the photoionized clouds of hundreds of megayears in the scenario where the undisrupted cooling flows feed the photoionized gas (see Section~\ref{sec:HI}). These lifetimes are comparable to the cooling time for \OVI-bearing gas of $100 (Z/0.3)^{-1} P_2^{-1}\,$Myr.  An argument in favor of the disruption happening at low temperatures is that photoionized clouds suffer from additional disruption mechanisms owing to hydrodynamical instabilities and conduction.  A counterargument is that the warmer gas may be more diffuse and hence more disruptable.  For example, it may be easier to channel the dynamical energy in sloshing motions toward reheating this warmer gas.

Lastly, the energetics to maintain a virialized atmosphere containing a fraction $f_g$ of a halo's associated gas is similar for $f_g\sim 1$ to that required to recycle cooling gas (as considered in eqn.~\ref{eqn:dotE}):
\begin{eqnarray}
\dot E &=& \frac{\Lambda M n}{\mu_i m_p}, \\
&=& 2\times10^{49} {\rm ~erg~ yr^{-1}} f_g P_2 T_6^{-1.7} \left( \frac{Z}{0.3}\right) \left(\frac{M_{\rm halo}}{10^{12}\Msun} \right), \nonumber
\end{eqnarray}
and the required $\dot E$ are increased by a factor of $1.4$ and $3.4$ for a lognormal distribution of temperatures centered at $T_6$ with standard deviations of $0.2$ and $0.4~$dex.

\paragraph{Conclusion} The large observed \OVI\ columns may owe to the virialized halo gas being distributed over a range of temperatures, with a factor of $\approx 2-3$ spread around $T_{\rm vir}$ and with most of the gas associated with an $L_*$ halo situated within the virial radius.  This virialized gas scenario likely requires similar fluxes of cooling gas through $T\sim 10^{5.5}$K as inferred in Section~\ref{sec:simple}; the primary difference with the cooling flow model presented in Section~\ref{ss:coolingflow} would be that the cooling gas is disrupted before reaching $T\approx 10^4$K.

\section{The relation of \OVI\ to the $\sim 10^4$K CGM}
\label{sec:HI}

We have thus far largely ignored the properties of the photoionized CGM ``clouds'' that show up as lower ionization (than \OVI) metal absorption systems.  These clouds are likely the end products of the \OVI-tracing cooling flows.  A robust constraint on the properties of the photoionized clouds can be derived from their \HI\ columns.  About half of the COS-Halos \HI\ absorbers are saturated in the observed Lyman series transitions, so their $N_{\rm HI}$ is only bounded from below.  However, the COS-Halos survey captures more Lyman-series transitions with increasing system redshift, allowing this survey to constrain $N_{\rm HI}$ to have a definite value for a significant fraction of their systems with $z=0.2-0.4$ \citep{tumlinsonCOS, prochaska17}.  Typically, they find $N_{\rm HI} \sim 10^{16}~{\rm cm^{-2}}$, albeit with scatter of $1~$dex below and $\sim 2~$dex above this value.  

From the measured $N_{\rm HI}$ and using the photoionization equilibrium relation $N_{\rm HI} = L_{\rm cold}^{\rm (HI)} \alpha n^2/\Gamma$,  where $\alpha = 4\times10^{-13}T_4^{-0.7}$cm$^{3}$\;s$^{-1}$ is the CASE A recombination coefficient and $\Gamma$ is the \HI\ photoionization rate, we can deduce the line-of-sight--integrated extent of the \HI\ absorbers:
\begin{equation}
L_{\rm cold}^{\rm (HI)} = 0.8~ {\rm kpc} ~~T_4^{2.7}  \Gamma_{-13} \left( \frac{\langle N_{\rm HI}P_{1}^{-2} \rangle}{10^{16} {\rm cm}^{-2}} \right),
\label{eqn:LHI}
\end{equation}
where we have placed the quantities that are most likely to vary in the average, $\langle ...\rangle$, and $\Gamma_{-13} \equiv \Gamma/10^{-13} {\rm \; s}^{-1}$.  Robust intergalactic \HI\ absorption modeling yields $\Gamma_{-13} = (1.0 \pm 0.3)\times [(1+z)/1.2]^5$ \citep{2016arXiv160502738G}, a functional form we henceforth refer to as $\Gamma^{\rm GKCS}(z)$; see also \citet{2015ApJ...811....3S} who measure a nearly identical form.\footnote{This estimate for $L_{\rm cold}^{\rm (HI)}$ does not apply at $N_{\rm HI}\gtrsim 10^{17.5}~{\rm ~cm}^{-2}$ owing to self-shielding, and at (potentially) small distances from the central galaxy, where proximate emission can be significant.  Proximate emission from the host galaxy is unlikely to affect the ionization at impact parameters of $\gtrsim 100~$kpc where many of the absorbers lie (Appendix~\ref{sec:proximity}).} Equation~(\ref{eqn:LHI}) shows that the \HI\ clouds should have sizes of $< 1~$kpc if they are in thermal pressure equilibrium with the pressure range that we inferred for the warmer gas, $P_{1} = 1-10$.  However, we show below that for many cold clouds, $P_{1} \lesssim (0.3-3)\times (Z/0.3)$, suggesting some nonthermal pressure support (or high metallicities of $Z\approx 1$).\footnote{Note that, since $L_{\rm cold}^{\rm (HI)}$ is the integrated line-of-sight extent, individual ``clouds'' could be much smaller in size such as advocated in \citet{mccourt16}.} Furthermore, the total mass in \HI-tracing gas with a radial extent $R$ can be estimated as
\begin{eqnarray}
M_{\rm cold}^{\rm (HI)} &\approx & \pi R^2 L_{\rm cold}^{\rm (HI)} \mu_i m_p n, \label{eqn:Mphoto}\\
&=& 3\times10^9\Msun  ~R^2_{200{\rm k}} T_4^{1.7}  \Gamma_{-13} \left( \frac{\langle N_{\rm HI} P_{1}^{-1} \rangle}{10^{16} {\rm cm}^{-2}} \right).\nonumber
\end{eqnarray}
Again using $P_{1} = 0.3-3 \times (Z/0.3)$ and $N_{\rm HI} = 10^{16}~{\rm cm^{-2}}$, results in $M_{\rm cold}^{\rm (HI)} = (3-30) \times10^9 (0.3/Z)\Msun$. 
 
 \begin{figure}
\begin{center}
\epsfig{file=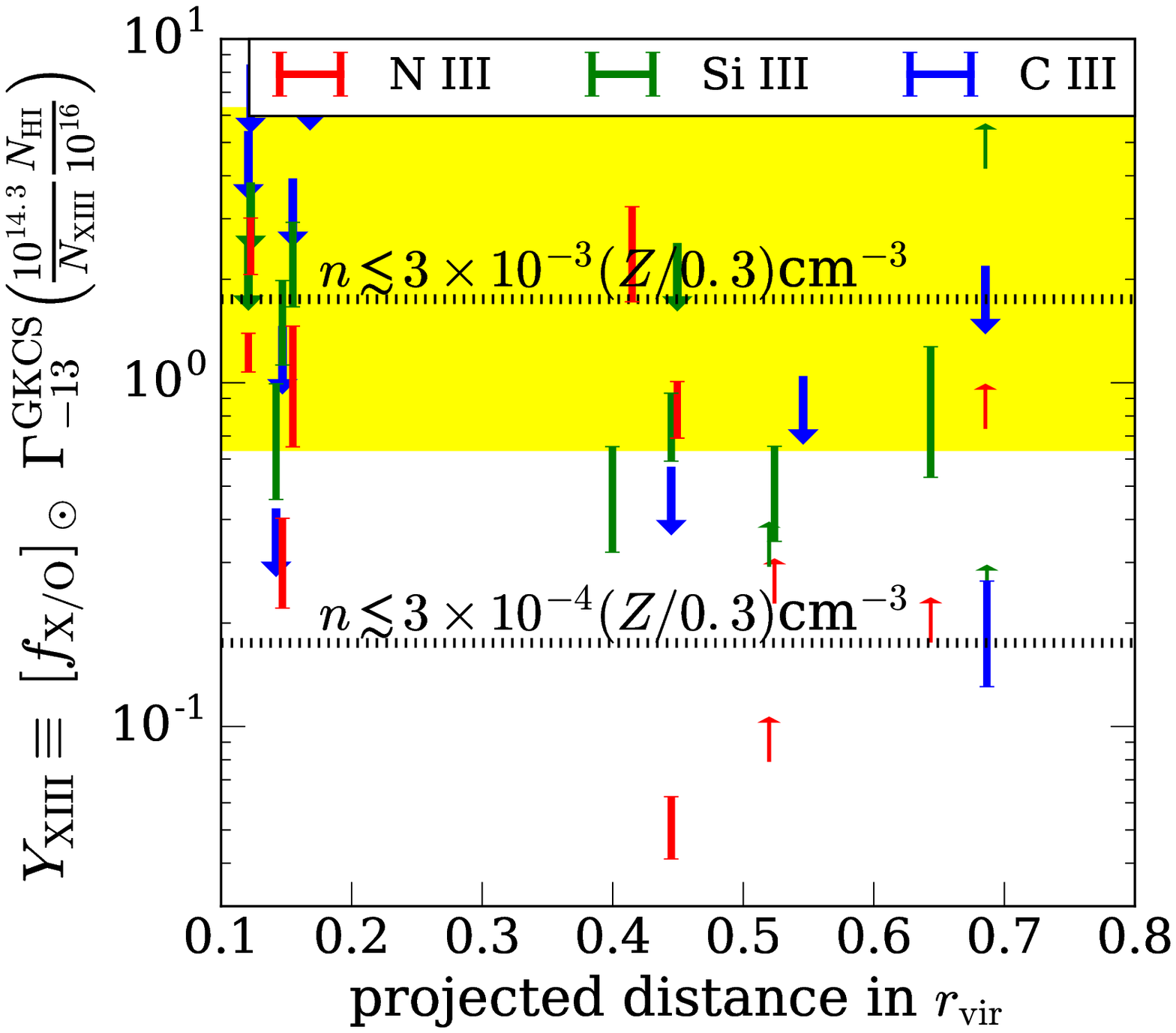, width=7cm}
\end{center}
\caption{The error bars show the estimated $Y_{\rm XIII}$ for all COS-Halos sightlines presented in \citet{werk14}, and supplemented with the follow-up in \citet{prochaska17}, for which $N_{\rm HI}$ has a well-constrained value with $N_{\rm HI}<10^{17.5}$cm$^{-2}$.   The upper bounds are absorbers with saturated $N_{\rm XIII}$, and the lower bounds are X{\sc iii} non-detections.  The error bars propagate the uncertainty in both column estimates. The horizontal lines give reference number densities for $Y_{\rm XIII}$ for the case where the absorber has solar metal abundance ratios, is exposed to the \citet{2016arXiv160502738G} ionizing background, has $T_4 = 1$, and $f_{\rm XIII} =1$. The yellow shaded region brackets the densities that we argue are required if we make the same assumptions as the horizontal lines and if the cold gas is in thermal pressure equilibrium with the warm and hot CGM gases over the range of pressures we argue is allowed, $P_1 = 1-10$.
 \label{fig:estimatedensity}}
\end{figure}

 \begin{figure}
\begin{center}
\epsfig{file=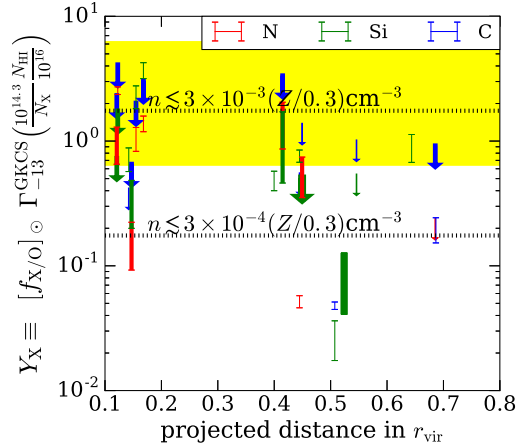, width=7cm}
\end{center}
\caption{The same as Figure~\ref{fig:estimatedensity2} except for $Y_{\rm X}$ rather than for $Y_{\rm XIII}$.  $Y_{\rm X}$ is defined on the $y$-axis, where all observed ions of an element are summed to estimate their $N_{\rm X}$, using C{\sc~ ii-iv}, N{\sc~ ii-iii}, or Si{\sc~ ii-iv}.   The line width of each error bar is proportional to the number of lines used in the estimate.  Unlike in Figure~\ref{fig:estimatedensity2}, lower bounds are not shown. 
 \label{fig:estimatedensity2}}
\end{figure}
 
  We can derive a constraint on the densities of CGM clouds by using our estimate for $M_{\rm cold}^{\rm (HI)}$ in conjunction with a mass estimate from twice-ionized metal absorption lines.  In particular, the average COS-Halos sightlines show \CIII\ with $N_{\rm CIII} \sim 10^{14.3}$cm$^{-2}$, \NIII\ with $N_{\rm NIII}\sim 10^{14.3}$cm$^{-2}$, and \SiIII\ with $N_{\rm SiIII} \sim 10^{13.5}$cm$^{-2}$, where these numbers are from crudely eyeballing the left panel in Figure~\ref{fig:columns}.  From these columns, we can estimate the mass that these ions trace by solving
\begin{equation}
\langle N_{\rm XIII} \rangle  = \overbrace{f_{\rm XIII} \,f_{\rm X/O} \, [f_{\rm O}]_\odot Z}^{\rm XIII ~per~ H} \times  \overbrace{\frac{0.75  M_{\rm cold}^{\rm (XIII)}}{\pi R^2 m_p}}^{\rm H~surface ~density}
\end{equation}
 for $M_{\rm cold}^{\rm (XIII)} $, where $0.75$ is the hydrogen mass fraction, $Z$ is the oxygen metallicity,  $f_{\rm XIII}$ is the fractional abundance of element X in state X~{\sc iii},  and $f_{\rm X/O}$ is the ratio of the number abundances in element X to oxygen.  In the Sun, $[f_{\rm X/O}]_{\odot}=  \{$0.7$, $0.2$, $0.06$\}$ for $X \in $\{C, N, Si\}.  The solution for $M_{\rm cold}^{\rm (XIII)} $ is
\begin{eqnarray}
M_{\rm cold}^{\rm (XIII)}                &=& 4.4\times 10^{8} \Msun~ f_{\rm X/O}^{-1} R_{100{\rm k}}^2 \label{eqn:MCIII}\\
                   &&\times\left(\frac{0.3}{f_{\rm XIII} Z}\right)  \left( \frac{ \langle N_{\rm XIII} \rangle}{10^{14.3}{\rm cm}^{-2}} \right). \nonumber
\end{eqnarray}

Because \HI\ and these doubly ionized metals are likely from the same gas, we can equate our expressions for $M_{\rm cold}^{\rm (HI)}$ and  $M_{\rm cold}^{\rm (XIII)}$ (eqn.s~\ref{eqn:Mphoto} and \ref{eqn:MCIII}), and solve for the density of photoionized gas:
\begin{equation}
n = 1.7\times10^{-3} {\rm \, cm}^{-3} \, T_4^{0.7}  \left(\frac{\Gamma}{\Gamma^{\rm GKCS}} \frac{f_{\rm X/O}}{[f_{\rm X/O}]_\odot}\right) \left(\frac{f_{\rm XIII} Z}{0.3}\right) Y_{\rm XIII}, \label{eqn:nest}
\end{equation}
where $Z$ is the oxygen metallicity relative to solar and
\begin{equation}
Y_{\rm XIII}\equiv [f_{\rm X/O}]_\odot \; \Gamma^{\rm GKCS}_{-13}(z) \left( \frac{10^{14.3}}{ N_{\rm XIII}} \frac{ N_{\rm HI}} {10^{16}}\right),
\label{eqn:Y}
\end{equation}
and $\Gamma^{\rm GKCS}_{-13}(z)$ accounts for the observed redshift evolution of the background.
Equation~(\ref{eqn:nest}) holds locally and, hence, we have removed the $\langle...\rangle$; it does not depend on the transverse extent of the \HI\ and metal ion absorbers.  

Since we expect $Y_{\rm XIII}\sim 1$ for \CIII\ and \NIII, as we have defined $Y_{\rm XIII}$ so that it is unity for roughly the average observed columns in these ions, equation~(\ref{eqn:nest}) suggests number densities of $n \sim 10^{-3} f_{\rm XIII} \; $cm$^{-3}$ and thermal pressures of $P\sim10 \,f_{\rm XIII}~$cm$^{-3}$K for the cold $10^4$K clouds, on the low side of what we argued was necessary for the warmer gas to explain the \OVI.  While an upper bound on the density results if we take $f_{\rm XIII}$'s maximum of unity, Appendix~\ref{sec:helpful} shows that the fractional abundance of $f_{\rm XIII}$ is near unity for the range of temperatures and pressures that our $Y_{\rm XIII}$ measurements suggest ($n\sim 10^{-4}-10^{-3}$cm$^{-3}$), even when considering non-equilibrium effects and different ionizing backgrounds.  The only exception is silicon, where $f_{\rm Si III}\approx 1$ at $\approx 10^{-3}$cm$^{-3}$, but this fraction declines quickly with decreasing density. 

The density and pressures can be directly estimated for individual systems by computing $Y_{\rm XIII} \propto N_{\rm HI}/N_{\rm XIII}$ from the measured columns in each system.  Figure~\ref{fig:estimatedensity} shows estimates of $Y_{\rm XIII}$ for all COS-Halos sightlines presented in \citet{werk13} and \citet{prochaska17} for which $N_{\rm HI}$ has a well-constrained value with $N_{\rm HI}<10^{17.5}$cm$^{-2}$. (The upper bounds are the case where the X{\sc iii} absorption is saturated and the lower bounds show unsaturated systems.)  The error bars propagate the uncertainty in both column estimates.  Note that, generally, $0.2< Y_{\rm XIII} <2$, suggesting densities of $3\times10^{-4} (Z/0.3) \lesssim n\lesssim 3\times10^{-3} (Z/0.3) $cm$^{-3}$ if $f_{\rm XIII} = 1$ (and lower densities if $f_{\rm XIII} < 1$), with an average of $Y_{\rm XIII}\approx 0.5$ and $n\approx 1\times10^{-3}(Z/0.3)$cm$^{-3}$.  These density estimates assume solar ratios for the metals such that ${f_{\rm X/O}}/{[f_{\rm X/O}]_\odot} = 1$ and also $\Gamma = \Gamma^{\rm GKCS}$.  The $Y_{\rm XIII}$ for silicon tends to lie somewhat above that for nitrogen and for carbon in the same system, noting that errors that align in radius are from the same sightline.  We think the higher values for silicon occur because, with decreasing density below $n= 10^{-3}$cm$^{-3}$, \SiIII\ has an ionic fraction that falls off faster than the other ions (see Appendix~\ref{sec:helpful}).\footnote{We note that the silicon is more likely to be depleted onto dust.  It is conservative to ignore depletion in our bounds as depletion pushes our estimates of $Y$ and $n$ higher, similar to the case where $f_{\rm XIII} <1$.}  We remark that these density inferences assume that the \HI\ traces the twice-ionized metals even though $N_{\rm HI} \propto n^2 T^{-0.7}$, whereas we argued that $N_{\rm XIII} \propto n$.  The upper bounds on the density of the metal line-tracing gas would be even lower if the \HI\ owes to denser embedded cloudlets.

 The potential weak points in our density constraints from $Y_{\rm XIII}$ are (1) we assumed that the metagalactic radiation background sets the ionization of the \HI\ and (2) we assumed that the \HI\ gas is photoionized with $T=10^4$K.  Regarding the first potential weak point, it may be the case for the crop of $Y$ measurements at $r \approx 0.15\,r_{\rm vir}$ (see Fig.~\ref{fig:estimatedensity}) that the local galaxy's emission dominates over the background.  However, local emission cannot dominate the flux for the absorbers at $\gtrsim 0.5\,r_{\rm vir}$, as discussed in Appendix~\ref{sec:proximity}.  Regarding the second potential weak point, somewhat higher temperatures would result in marginally larger $n$ as $n \propto T^{0.7}$.  However, our estimates would err significantly if $T$ were sufficiently large that the \HI\ becomes collisionally ionized.  The \HI\ and metals cannot be from hotter gas in the \OVI--producing cooling flow as the columns of ions are an order of magnitude too small (see Section~\ref{ss:coolingflow} and Fig.~\ref{fig:TI}).  Instead, higher temperature gas would have to arise from more thermally stable gas that experiences additional heating beyond photoionization.  Such heating would have to be substantial since maintaining a temperature of $T_{4.5}$ requires
 \begin{eqnarray}
 \dot E &=& \frac{M_{\rm cold}^{\rm (HI)}}{\mu_i m_p} \frac{5 k_b T}{t_{\rm cool}(T)}, \\
            &\approx& 5\times10^{48} {\rm ~erg ~yr^{-1}~}\times R^2_{200{\rm k}} \,T_{4.5}^{2.7} \, \Gamma_{-13} \nonumber \\
            && \times \left( \frac{N_{\rm HI}}{10^{16} {\rm cm}^{-2}} \right)\left( \frac{Z}{0.3} \right).  \label{eqn:heatcold}
 \end{eqnarray}
 The second line uses that $t_{\rm cool}(T) \approx 10^2 P_{1}^{-1} (Z/0.3)^{-1}$Myr, valid for $2\times10^4 < T < 1\times10^5$K (Fig.~\ref{fig:tcool} in Appendix~\ref{sec:helpful}).   The inferred heating from equation~(\ref{eqn:heatcold}) is a factor of $\sim 10\times T_{4.5}^{-2.7}$ smaller than that required to maintain the warm and hot gas (see eqn.~\ref{eqn:dotE}).  We leave as an open question whether such significant (and seemingly fine-tuned) heat dissipation can occur relatively uniformly across the photoionized ``clouds''.\footnote{A large potential source of heating is conduction from the hot atmosphere:  The mean free path is generally larger than the estimated size of the clouds, and, in the limit of saturated conduction with flux of hot gas $5 n k_b T c_s$ \citep{cowie77}, we find heating timescales of $\sim L_{\rm cloud}/$[1~kpc] Myr.  However, magnetic fields can strongly suppress conduction \citep{spitzer65} and there is strong evidence that they do \citep[e.g.][]{mccourt16}.  Also, we note that to substantially change our conclusions regarding $n$, the gas must be ``stable'' at a temperature where the \HI\ is collisionally ionized ($\gtrsim 5\times10^4$K) but where the twice-ionized metals are still not ($\lesssim 10^5$K).  Otherwise, the dependence of our density estimate on temperature is weak ($n\propto T^{0.7}$).} 

A tighter limit on $n$ derives from, rather than considering twice-ionized species as in $Y_{\rm XIII}$, summing all observed ions for element $X$ along a sightline to estimate $N_{\rm X}$ and replacing $N_{\rm XIII}$ with $N_{\rm X}$ in our expression for $Y_{\rm XIII}$.  We call the new quantity $Y_{\rm X}$.  The derivation for our formula regarding X{\sc ~iii} also applies to X.  Figure~\ref{fig:estimatedensity2} shows $Y_{\rm X}$, summing when available C{\sc~ ii-iv}, N{\sc ~ii-iii}, or Si{\sc ~ii-iv} to calculate $N_{\rm X}$, with the line width proportional to the number of ions used in the estimate.  The estimates are marginally lower than for $Y_{\rm XIII}$, but generally consistent. The largest changes when going from $Y_{\rm XIII}$ to $Y_{\rm X}$ are the two lowest $Y_{\rm X}$ estimates from silicon, where the \SiIV\ column dominates (\SiIV\ being the primary ionization stage expected at the low densities that these $Y_{\rm X}$ suggest).  The small shifts for most values going from $Y_{\rm XIII}$ to $Y_{\rm X}$ are consistent with the expectation that the twice-ionized species is most present for $n \sim 10^{-3}$cm$^{-3}$. 

To conclude this section, the upper bounds on the density from taking the ratio $N_{\rm HI}/N_{\rm XIII}$ and $N_{\rm HI}/N_{\rm X}$ are robust.  They imply that the densities of some low-ionization CGM absorbers are somewhat below the CGM pressures we require for \OVI-tracing gas.  The low densities suggest that these ``clouds'' are nonthermally supported by cosmic rays and/or magnetic fields, although we note alternatively that metallicities closer to solar alleviate some tension with the pressure range needed for thermal pressure equilibrium.  The $Y_{\rm X}$ of a few clouds suggest densities of $n <10^{-4} (Z/0.3)$cm$^{-3}$, well out of pressure equilibrium without supersolar metallicities. Taking the mean inference of $Y_{\rm XIII}\approx 0.5$ such that $n < 10^{-3}(Z/0.3)$cm$^{-3}$, the implied gas masses in photoionized clouds are $M_{\rm cold}> 3\times10^{9} (0.3/Z)\Msun$.  However, we caution that the total mass is not proportional to $\langle Y \rangle^{-1}$ but instead to $\langle Y^{-1} \rangle$ and so the lowest density clouds dominate $M_{\rm cold}$.  Thus, our $M_{\rm cold}$ estimate based on the median $Y$ is likely to be an underestimate.  Assuming our bound on $M_{\rm cold}$ is saturated and that the \OVI\ owes to cooling flows, the $\dot M$ we infer from the \OVI\ is sufficient to re-form our estimated $M_{\rm cold}$ in $(0.03-0.3)\times (0.3/Z)~$Gyr for our preferred range of $P=10-100\; $cm$^{-3}$K for the warm gas, shorter than the halo dynamical time at distances of $\sim r_{\rm vir}$ of $\sim 1\,$Gyr.\footnote{We note that the mean metallicity is constrained to be $Z=0.3$ from photoionization models, although there is more than an order of magnitude scatter in $Z$ in this modeling with $Z = 0.01-3$ (and a curious $Z-N_{\rm HI}$ anticorrelation; \citealt{prochaska17}).  This large scatter, if correct, would result in a single density producing a wide range of $Y_{\rm XIII}$, complicating the density estimates in this section.}    Cloud survival times of tens to hundreds of millions of years are in accord with rough expectations for disruption timescales from hydrodynamical instabilities (Appendix~\ref{sec:kinematics}).  Thermal conduction could also evaporate these clouds within such times.  Forming clouds in less than a dynamical time could explain why these clouds appear to be entrained with the rest of the halo gas, as evidenced by their kinematic coincidence with the \OVI. 

Low densities of $n=10^{-5}-10^{-3}$cm$^{-3}$ for the photoionized clouds have been estimated from slab photoionization models \citep{stocke13, werk14}.  Here, we found densities that are in the upper half of this density range if $Z\sim 0.3$.  \citet{werk14} estimated the density of not just the absorbers with well-constrained $N_{\rm HI}$ as done here, but also absorbers with only a lower bound on $N_{\rm HI}$.  For the systems with only lower bounds, \citet{werk14} selected a definite value for $N_{\rm HI}$ based on various considerations to use in their photoionization modeling.  Equation~(\ref{eqn:nest}) suggests that choosing $N_{\rm HI}$ values that are too low would underestimate $n$.

\section{how our models relate to CGM simulations and analytic models}
\label{sec:simulations}
 Here we comment on how the results derived from previous analytic models and numerical simulations of the CGM relate to the picture of the CGM constructed here. 

{
\citet{heckman02} mirrors many of the considerations here, using arguments based on density and energetics to constrain the spatial extent of \OVI\ absorbers seen in both Galactic and intergalactic contexts.  A major distinction between this study and theirs is that \citet{heckman02} focused on the ubiquity of processes that produce $N_{\rm OVI}\sim 10^{14.5}~$cm$^{-2}$. We have used that this characteristic column, combined with the observed extent of CGM \OVI, to limit the density to be $P_2 <1$ on total energy grounds.  Such low densities require the physical sizes of the regions producing the \OVI\ to be $>30$kpc, too large to be boundary layers around the (we estimate) $\sim 1\,$kpc $10^4$K clouds or shocks within these clouds. We disagree with \citet{heckman02} that, for the case of large-scale cooling gas, this should lead to a characteristic column density.  Rather, we argue that the observed values for $N_{\rm OVI}$ are driven by the projected mass flux through a given temperature, $\sim \dot M/r_{\rm vir}^2$, in $L_*$-galaxy halos.}

{\citet{oppenheimer16} argued that the CGM \OVI\ comes from virial temperature gas, supported by their observations of \OVI\ in  cosmological simulations.  We have argued that the large spread of inferred halo masses in the COS-Halos sample requires a comparable spread in the temperatures of gas in each halo in order to explain the uniformity of  \OVI\ columns.   If there is a distribution of temperatures, a dynamic multiphase picture becomes more likely because of the strong dependence of the cooling time on temperature.   
   A larger fraction of the \OVI-bearing gas in the \citet{oppenheimer16} simulations may be sourced by cooling from recently accreted material because they find columns that are a few times below those observed (Section~\ref{sec:simple}).
}

\citet{fielding16} ran idealized simulations of CGM gas by following the spherical collapse of an overdense region and then by injecting energy at an inner boundary in response to accretion to mimic galactic feedback.  The densities of virial temperature gas in their halos lie mostly within the range we have argued for, with \citet{fielding16} finding $P_2 \approx 3$ at $0.1 ~r_{\rm vir}$ and $P_2\approx 0.1$ at $1 ~r_{\rm vir}$.  In addition, assuming $Z=0.3 \, Z_\odot$, they found a factor of a few smaller $N_{\rm OVI}$ than is observed at $10^{12}\Msun$, with $N_{\rm OVI}$ depending weakly on halo mass, and with a weak dependence on $R$.  \citet{fielding16} chose to inject approximately all of the energy available from supernova feedback into their simulations' CGMs.  Such energetic feedback is consistent with the energetics we estimate are needed to maintain the \OVI\ (and the larger observed columns than in the \citealt{fielding16} simulations suggest even larger feedback energies are needed).  The \citet{fielding16} simulations generate negligible \HI\ absorption, with $N_{\rm HI} <10^{14}$cm$^{-2}$ at impact parameters of  $R \sim 100\;$kpc.  This lack of \HI\ suggests that their simulated flows of cooling gas are largely reheated before reaching $\sim10^5$K, although it may alternatively owe to insufficient resolution.  

The simulations presented in \citet{salem16} have been the most successful to date at generating a CGM that replicates ionic column densities typical of the observed CGM of $L_*$ galaxies.  Like studies before them, \citet{salem16} re-simulated zoom-in regions around halos in a larger cosmological simulation.  However, a major difference with many previous works was that they included cosmic ray feedback.  All of their runs assumed $30\%$ of the supernova energy went into cosmic rays (which is unlikely to be dissipated near the injection sight) and that $70\%$ went into local heating (where it is less clear where the energy is dissipated).  Cosmic rays often dominate the gas pressure in their simulations, in some places exceeding thermal pressure by a factor of $10$.   This nonthermal pressure support is likely why the \citet{salem16} simulations fare better at reproducing the observed columns of low ions; we argued that gas in thermal pressure equilibrium with the hot atmosphere is unlikely to generate sufficient columns of certain twice-ionized metals.

\section{conclusions}
\label{sec:conclusions}
Little is known about the circumgalactic media of $L_*$ galaxies, including how much gas resides in them, how this gas is distributed over thermal phases, and how the gas accretes onto central galaxies.  Through simple estimates compared to HST/COS observations, we have attempted to develop a picture for the CGM, with the goal of addressing these outstanding issues.  Our particular focus has been on the large columns of $N_{\rm OVI} \sim 10^{14.5}$cm$^{-2}$ that extend to the halo virial radius.  We argued that the \OVI\ cannot be generated in the vast quantities required by turbulent boundary layers, by shocks penetrating the $\sim10^4$K clouds, or by hot winds, in addition to presenting other arguments against these scenarios.  Additionally, we showed that if the \OVI\ owes to photoionized gas in thermal equilibrium, this scenario requires extremely low gas densities of $n \lesssim 10^{-5} {\rm cm}^{-3}$, nonthermal pressure support that is $10-100\times$ thermal pressure, and ``cloud'' sizes that are a significant fraction of the virial radius -- a scenario that seems unlikely.

Instead, we argued that the \OVI\ around $L_*$ galaxies likely owes to massive flows of cooling gas.  This scenario requires mass fluxes through $T\sim 10^{5.5}$K of $\dot M = 10 P_1 \Msun~$yr$^{-1}$ to explain the observed $N_{\rm OVI}$ with the bounds $1 < P_1 <10$.  The upper bound is set by the maximum pressure that is energetically feasible, and the lower bound is set by cooling and total mass considerations.  An ${\cal O}(1)$ fraction of all the gas associated with each $\sim 10^{12}\Msun$ halo has to be cooling to generate these mass fluxes.  The reduced incidence of \OVI\ in quiescent galaxies requires them to have smaller $\dot M$ in this scenario. 
 
 The other viable picture for the \OVI\ is that, rather than a flow from hot to cold, the \OVI\ owes to gas that stays relatively hot.  The large \OVI\ columns cannot be from the simplistic picture of all gas at a temperature of $T_{\rm vir}$, as the observed values of $N_{\rm OVI}$ are generally too high and show too weak of a dependence on halo mass.  However, a distribution of temperatures around $T_{\rm vir}$ with half of a decade standard deviation could explain these dependences.  Since the colder gas in this distribution has much shorter cooling times, this gas is likely cooling and condensing, and to generate the \OVI\ requires similar $\dot M$ through the \OVI-bearing temperatures of $\sim 10^{5.5}$K as in our cooling gas model.  However, in this picture, feedback is reheating the gas before it settles to $\sim 10^4$K.  We discussed the pros and cons of both viable pictures, and reality may be a combination of both.  One reason for preferring the cooling gas scenario is the existence of cold clouds (which would have to be much longer lived, longer than a halo dynamical time if the flow to $10^4$K is reduced by an order of magnitude).  
 
 Both viable models require the feedback energy input into the $z\sim0$ CGM of $L_*$ galaxies to be a significant fraction of the total kinetic energy in galactic supernovae and stellar winds.  Feedback models for which the energetics is suppressed by an order of magnitude are disfavored, such as those where feedback is dominated by cosmic rays, unclustered supernovae, or momentum.  
  
 In addition, we discussed the properties of lower ionization clouds that are predominantly photoionized.  The densities of these clouds can be estimated using that the column of neutral hydrogen scales as $N_{\rm HI} \propto n^2$ and that the column density in ion X{\sc iii} scales as $N_{\rm XIII} \propto f_{\rm XIII}\, n$.  Taking $f_{\rm XIII} < 1$ gives an upper bound on the density of $n< (0.1-3)\times 10^{-3} (Z/0.3) {\rm cm^{-3}}$, and we argued that $f_{\rm XIII} \approx 1$ for X~$\in$~\{C, N, Si\} so that this bound is likely to be near saturation.  The pressures of these clouds are lower than those that were required to explain the warmer \OVI-bearing gas.  This qualitatively confirms the low (but not the lowest) densities inferred from detailed ionization models \citep{stocke13, werk14}, and the low thermal pressures suggest that these clouds are supported by non-thermal pressure.  

The near-unity covering fraction of CGM absorbers, their clustered kinematics, and the necessity of $\sim 10^{11}\Msun$ reservoirs to explain the \OVI\ all suggest global sloshing of the intrahalo gas rather than independent ``clouds''.  The sloshing motions would have to be intense, as in projection most absorbers have velocities of $50-150\,$km~s$^{-1}$.  In this picture, the kinematic association of lower ionization ``clouds'' with \OVI-bearing ones owes to both types comoving with the turbulent virialized atmosphere, with the \OVI\ regions also tracing gas farther out in the halo and hence having somewhat broader line widths.  The inferred $\dot M$ into cold clouds in the cooling flow scenario is able to create the cold clouds in a fraction of a halo dynamical time.  Therefore, buoyancy may not have enough time to accelerate warm and cold clouds relative to the hot atmosphere (and, in some viable parameter space, the terminal velocity of the clouds is also rather low).

We hope that the simple considerations outlined in this paper are tested and refined with simulations and against additional observations.  Our inferences motivate running CGM simulations of thermal instability as well as ones that include sources of nonthermal pressure support.  Our calculations also suggest that global simulations of the CGM are required to understand the observed \OVI.  Lastly, there is room to improve the line-ratio diagnostics and simple ionization modeling performed here as a complementary method to full ionization modeling. \\

\acknowledgments

MM thanks Drummond Fielding, Avery Meiksin, Peng Oh, and Jonathan Stern for helpful conversations.  We thank J. Xavier Prochaska, Jonathan Stern, and the anonymous referee for helpful comments on the manuscript.  MM thanks the Institute for Advanced Study visiting faculty program and the John Bahcall fellowship for support during much of this project, and the Aspen Center for Physics (NSF grant PHY-1066293), where this study was initiated.  This work is supported by an award from the Royalty Research Fund at the University of Washington, by NASA through the Space Telescope
Science Institute grant HST-AR-14307, by NSF through award AST-1614439, and by the Alfred P. Sloan Foundation.  

\bibliography{References}

\begin{thebibliography}{}

\bibitem[\protect\citeauthoryear{{Asplund} et~al.}{{Asplund}
  et~al.}{2009}]{2009ARA&A..47..481A}
{Asplund}, M., {Grevesse}, N., {Sauval}, A.~J.,  \& {Scott}, P. 2009, \araa,
  47, 481

\bibitem[\protect\citeauthoryear{{Barkana} \& {Loeb}}{{Barkana} \&
  {Loeb}}{2001}]{barkana01}
{Barkana}, R.,  \& {Loeb}, A. 2001, \physrep, 349, 125

\bibitem[\protect\citeauthoryear{{Begelman} \& {Fabian}}{{Begelman} \&
  {Fabian}}{1990}]{begelman90}
{Begelman}, M.~C.,  \& {Fabian}, A.~C. 1990, \mnras, 244, 26P

\bibitem[\protect\citeauthoryear{{Behroozi}, {Conroy}, \&
  {Wechsler}}{{Behroozi} et~al.}{2010}]{2010ApJ...717..379B}
{Behroozi}, P.~S., {Conroy}, C.,  \& {Wechsler}, R.~H. 2010, \apj, 717, 379

\bibitem[\protect\citeauthoryear{{Bordoloi} et~al.}{{Bordoloi}
  et~al.}{2017}]{2016arXiv160507187B}
{Bordoloi}, R., {Wagner}, A.~Y., {Heckman}, T.~M.,  \& {Norman}, C.~A. 2017,
  \apj, 848, 122

\bibitem[\protect\citeauthoryear{{Cantalupo}}{{Cantalupo}}{2010}]{cantalupo10}
{Cantalupo}, S. 2010, \mnras, 403, L16

\bibitem[\protect\citeauthoryear{{Chevalier} \& {Clegg}}{{Chevalier} \&
  {Clegg}}{1985}]{1985Natur.317...44C}
{Chevalier}, R.~A.,  \& {Clegg}, A.~W. 1985, \nat, 317, 44

\bibitem[\protect\citeauthoryear{{Cowie} \& {McKee}}{{Cowie} \&
  {McKee}}{1977}]{cowie77}
{Cowie}, L.~L.,  \& {McKee}, C.~F. 1977, \apj, 211, 135

\bibitem[\protect\citeauthoryear{{Crawford} et~al.}{{Crawford}
  et~al.}{2001}]{2001ApJ...553..367C}
{Crawford}, F., {Kaspi}, V.~M., {Manchester}, R.~N., {Lyne}, A.~G., {Camilo},
  F.,  \& {D'Amico}, N. 2001, \apj, 553, 367

\bibitem[\protect\citeauthoryear{{Draine} \& {Salpeter}}{{Draine} \&
  {Salpeter}}{1979}]{1979ApJ...231...77D}
{Draine}, B.~T.,  \& {Salpeter}, E.~E. 1979, \apj, 231, 77

\bibitem[\protect\citeauthoryear{{Faerman}, {Sternberg}, \& {McKee}}{{Faerman}
  et~al.}{2017}]{faerman16}
{Faerman}, Y., {Sternberg}, A.,  \& {McKee}, C.~F. 2017, \apj, 835, 52

\bibitem[\protect\citeauthoryear{{Fang}, {Bullock}, \& {Boylan-Kolchin}}{{Fang}
  et~al.}{2013}]{fang13}
{Fang}, T., {Bullock}, J.,  \& {Boylan-Kolchin}, M. 2013, \apj, 762, 20

\bibitem[\protect\citeauthoryear{{Faucher-Gigu{\`e}re}, {Kere{\v s}}, \&
  {Ma}}{{Faucher-Gigu{\`e}re} et~al.}{2011}]{2011MNRAS.417.2982F}
{Faucher-Gigu{\`e}re}, C.-A., {Kere{\v s}}, D.,  \& {Ma}, C.-P. 2011, \mnras,
  417, 2982

\bibitem[\protect\citeauthoryear{{Feldmann}, {Hooper}, \& {Gnedin}}{{Feldmann}
  et~al.}{2013}]{feldmann13}
{Feldmann}, R., {Hooper}, D.,  \& {Gnedin}, N.~Y. 2013, \apj, 763, 21

\bibitem[\protect\citeauthoryear{{Fielding} et~al.}{{Fielding}
  et~al.}{2017}]{fielding16}
{Fielding}, D., {Quataert}, E., {McCourt}, M.,  \& {Thompson}, T.~A. 2017,
  \mnras, 466, 3810

\bibitem[\protect\citeauthoryear{{Ford} et~al.}{{Ford} et~al.}{2016}]{ford15}
{Ford}, A.~B., et~al. 2016, \mnras, 459, 1745

\bibitem[\protect\citeauthoryear{{Fox} et~al.}{{Fox} et~al.}{2005}]{fox05}
{Fox}, A.~J., {Wakker}, B.~P., {Savage}, B.~D., {Tripp}, T.~M., {Sembach},
  K.~R.,  \& {Bland-Hawthorn}, J. 2005, \apj, 630, 332

\bibitem[\protect\citeauthoryear{{Gaikwad} et~al.}{{Gaikwad}
  et~al.}{2017}]{2016arXiv160502738G}
{Gaikwad}, P., {Khaire}, V., {Choudhury}, T.~R.,  \& {Srianand}, R. 2017,
  \mnras, 466, 838

\bibitem[\protect\citeauthoryear{{Gaspari}, {Temi}, \& {Brighenti}}{{Gaspari}
  et~al.}{2017}]{2017MNRAS.466..677G}
{Gaspari}, M., {Temi}, P.,  \& {Brighenti}, F. 2017, \mnras, 466, 677

\bibitem[\protect\citeauthoryear{{Giacintucci} et~al.}{{Giacintucci}
  et~al.}{2014}]{2014ApJ...781....9G}
{Giacintucci}, S., {Markevitch}, M., {Venturi}, T., {Clarke}, T.~E., {Cassano},
  R.,  \& {Mazzotta}, P. 2014, \apj, 781, 9

\bibitem[\protect\citeauthoryear{{Gnat} \& {Sternberg}}{{Gnat} \&
  {Sternberg}}{2007}]{2007ApJS..168..213G}
{Gnat}, O.,  \& {Sternberg}, A. 2007, \apjs, 168, 213

\bibitem[\protect\citeauthoryear{{Gnat} \& {Sternberg}}{{Gnat} \&
  {Sternberg}}{2009}]{2009ApJ...693.1514G}
{Gnat}, O.,  \& {Sternberg}, A. 2009, \apj, 693, 1514

\bibitem[\protect\citeauthoryear{{Green} et~al.}{{Green}
  et~al.}{2012}]{green12}
{Green}, J.~C., et~al. 2012, \apj, 744, 60

\bibitem[\protect\citeauthoryear{{Gupta} et~al.}{{Gupta}
  et~al.}{2012}]{gupta12}
{Gupta}, A., {Mathur}, S., {Krongold}, Y., {Nicastro}, F.,  \& {Galeazzi}, M.
  2012, \apjl, 756, L8

\bibitem[\protect\citeauthoryear{{Haardt} \& {Madau}}{{Haardt} \&
  {Madau}}{2001}]{haardt01}
{Haardt}, F.,  \& {Madau}, P. 2001, in Clusters of Galaxies and the High
  Redshift Universe Observed in X-rays, ed. D.~M. {Neumann} \& J.~T.~V. {Tran},
  64

\bibitem[\protect\citeauthoryear{{Haardt} \& {Madau}}{{Haardt} \&
  {Madau}}{2012}]{haardt12}
{Haardt}, F.,  \& {Madau}, P. 2012, \apj, 746, 125

\bibitem[\protect\citeauthoryear{{Heckman} et~al.}{{Heckman}
  et~al.}{2002}]{heckman02}
{Heckman}, T.~M., {Norman}, C.~A., {Strickland}, D.~K.,  \& {Sembach}, K.~R.
  2002, \apj, 577, 691

\bibitem[\protect\citeauthoryear{{Hopkins} \& {Beacom}}{{Hopkins} \&
  {Beacom}}{2006}]{hopkinsbeacom}
{Hopkins}, A.~M.,  \& {Beacom}, J.~F. 2006, \apj, 651, 142

\bibitem[\protect\citeauthoryear{{Hummels} et~al.}{{Hummels}
  et~al.}{2013}]{hummels13}
{Hummels}, C.~B., {Bryan}, G.~L., {Smith}, B.~D.,  \& {Turk}, M.~J. 2013,
  \mnras, 430, 1548

\bibitem[\protect\citeauthoryear{{Johnson}, {Chen}, \& {Mulchaey}}{{Johnson}
  et~al.}{2015}]{johnson15}
{Johnson}, S.~D., {Chen}, H.-W.,  \& {Mulchaey}, J.~S. 2015, \mnras, 449, 3263

\bibitem[\protect\citeauthoryear{{Khaire} \& {Srianand}}{{Khaire} \&
  {Srianand}}{2015}]{2015MNRAS.451L..30K}
{Khaire}, V.,  \& {Srianand}, R. 2015, \mnras, 451, L30

\bibitem[\protect\citeauthoryear{{Kwak} \& {Shelton}}{{Kwak} \&
  {Shelton}}{2010}]{kwak10}
{Kwak}, K.,  \& {Shelton}, R.~L. 2010, \apj, 719, 523

\bibitem[\protect\citeauthoryear{{Lehner} \& {Howk}}{{Lehner} \&
  {Howk}}{2011}]{2011Sci...334..955L}
{Lehner}, N.,  \& {Howk}, J.~C. 2011, Science, 334, 955

\bibitem[\protect\citeauthoryear{{Leitherer} et~al.}{{Leitherer}
  et~al.}{1999}]{1999ApJS..123....3L}
{Leitherer}, C., et~al. 1999, \apjs, 123, 3

\bibitem[\protect\citeauthoryear{{Maller} \& {Bullock}}{{Maller} \&
  {Bullock}}{2004}]{maller04}
{Maller}, A.~H.,  \& {Bullock}, J.~S. 2004, \mnras, 355, 694

\bibitem[\protect\citeauthoryear{{Manchester} et~al.}{{Manchester}
  et~al.}{2006}]{2006ApJ...649..235M}
{Manchester}, R.~N., {Fan}, G., {Lyne}, A.~G., {Kaspi}, V.~M.,  \& {Crawford},
  F. 2006, \apj, 649, 235

\bibitem[\protect\citeauthoryear{{McBride}, {Fakhouri}, \& {Ma}}{{McBride}
  et~al.}{2009}]{mcbride09}
{McBride}, J., {Fakhouri}, O.,  \& {Ma}, C.-P. 2009, \mnras, 398, 1858

\bibitem[\protect\citeauthoryear{{McCourt} et~al.}{{McCourt}
  et~al.}{2016}]{mccourt16}
{McCourt}, M., {Oh}, S.~P., {O'Leary}, R.~M.,  \& {Madigan}, A.-M. 2016,
  ArXiv:1610.01164

\bibitem[\protect\citeauthoryear{{McCourt} et~al.}{{McCourt}
  et~al.}{2015}]{mccourt15}
{McCourt}, M., {O'Leary}, R.~M., {Madigan}, A.-M.,  \& {Quataert}, E. 2015,
  \mnras, 449, 2

\bibitem[\protect\citeauthoryear{{McCourt} et~al.}{{McCourt}
  et~al.}{2012}]{mccourt12}
{McCourt}, M., {Sharma}, P., {Quataert}, E.,  \& {Parrish}, I.~J. 2012, \mnras,
  419, 3319

\bibitem[\protect\citeauthoryear{{McQuinn}}{{McQuinn}}{2014}]{2014ApJ...780L..33M}
{McQuinn}, M. 2014, \apjl, 780, L33

\bibitem[\protect\citeauthoryear{{Meiksin}}{{Meiksin}}{2016}]{meskin16}
{Meiksin}, A. 2016, \mnras, 461, 2762

\bibitem[\protect\citeauthoryear{{M{\'e}nard} et~al.}{{M{\'e}nard}
  et~al.}{2010}]{menard10}
{M{\'e}nard}, B., {Scranton}, R., {Fukugita}, M.,  \& {Richards}, G. 2010,
  \mnras, 405, 1025

\bibitem[\protect\citeauthoryear{{Miller} \& {Bregman}}{{Miller} \&
  {Bregman}}{2015}]{miller15}
{Miller}, M.~J.,  \& {Bregman}, J.~N. 2015, \apj, 800, 14

\bibitem[\protect\citeauthoryear{{Mo} \& {Miralda-Escude}}{{Mo} \&
  {Miralda-Escude}}{1996}]{mo96}
{Mo}, H.~J.,  \& {Miralda-Escude}, J. 1996, \apj, 469, 589

\bibitem[\protect\citeauthoryear{{Moster} et~al.}{{Moster}
  et~al.}{2010}]{2010ApJ...710..903M}
{Moster}, B.~P., {Somerville}, R.~S., {Maulbetsch}, C., {van den Bosch}, F.~C.,
  {Macci{\`o}}, A.~V., {Naab}, T.,  \& {Oser}, L. 2010, \apj, 710, 903

\bibitem[\protect\citeauthoryear{{Murray}, {Quataert}, \& {Thompson}}{{Murray}
  et~al.}{2005}]{2005ApJ...618..569M}
{Murray}, N., {Quataert}, E.,  \& {Thompson}, T.~A. 2005, \apj, 618, 569

\bibitem[\protect\citeauthoryear{{Oppenheimer} et~al.}{{Oppenheimer}
  et~al.}{2016}]{oppenheimer16}
{Oppenheimer}, B.~D., et~al. 2016, \mnras, 460, 2157

\bibitem[\protect\citeauthoryear{{Oppenheimer} \& {Schaye}}{{Oppenheimer} \&
  {Schaye}}{2013a}]{2013MNRAS.434.1063O}
{Oppenheimer}, B.~D.,  \& {Schaye}, J. 2013a, \mnras, 434, 1063

\bibitem[\protect\citeauthoryear{{Oppenheimer} \& {Schaye}}{{Oppenheimer} \&
  {Schaye}}{2013b}]{oppenheimer13}
{Oppenheimer}, B.~D.,  \& {Schaye}, J. 2013b, \mnras, 434, 1043

\bibitem[\protect\citeauthoryear{{Oppenheimer} et~al.}{{Oppenheimer}
  et~al.}{2017}]{2017arXiv170507897O}
{Oppenheimer}, B.~D., {Segers}, M., {Schaye}, J., {Richings}, A.~J.,  \&
  {Crain}, R.~A. 2017, ArXiv:1705.07897

\bibitem[\protect\citeauthoryear{{Peek}, {M{\'e}nard}, \& {Corrales}}{{Peek}
  et~al.}{2015}]{2015ApJ...813....7P}
{Peek}, J.~E.~G., {M{\'e}nard}, B.,  \& {Corrales}, L. 2015, \apj, 813, 7

\bibitem[\protect\citeauthoryear{{Peeples} et~al.}{{Peeples}
  et~al.}{2014}]{peeples14}
{Peeples}, M.~S., {Werk}, J.~K., {Tumlinson}, J., {Oppenheimer}, B.~D.,
  {Prochaska}, J.~X., {Katz}, N.,  \& {Weinberg}, D.~H. 2014, \apj, 786, 54

\bibitem[\protect\citeauthoryear{{Prochaska} et~al.}{{Prochaska}
  et~al.}{2011}]{2011ApJ...740...91P}
{Prochaska}, J.~X., {Weiner}, B., {Chen}, H.-W., {Mulchaey}, J.,  \& {Cooksey},
  K. 2011, \apj, 740, 91

\bibitem[\protect\citeauthoryear{{Prochaska} et~al.}{{Prochaska}
  et~al.}{2017}]{prochaska17}
{Prochaska}, J.~X., et~al. 2017, \apj, 837, 169

\bibitem[\protect\citeauthoryear{{Richter} et~al.}{{Richter}
  et~al.}{2017}]{richter17}
{Richter}, P., et~al. 2017, \aap, 607, A48

\bibitem[\protect\citeauthoryear{{Salem} et~al.}{{Salem}
  et~al.}{2015}]{2015ApJ...815...77S}
{Salem}, M., {Besla}, G., {Bryan}, G., {Putman}, M., {van der Marel}, R.~P.,
  \& {Tonnesen}, S. 2015, \apj, 815, 77

\bibitem[\protect\citeauthoryear{{Salem}, {Bryan}, \& {Corlies}}{{Salem}
  et~al.}{2016}]{salem16}
{Salem}, M., {Bryan}, G.~L.,  \& {Corlies}, L. 2016, \mnras, 456, 582

\bibitem[\protect\citeauthoryear{{Shapiro} \& {Benjamin}}{{Shapiro} \&
  {Benjamin}}{1991}]{1991PASP..103..923S}
{Shapiro}, P.~R.,  \& {Benjamin}, R.~A. 1991, \pasp, 103, 923

\bibitem[\protect\citeauthoryear{{Sharma} et~al.}{{Sharma}
  et~al.}{2012}]{sharma12b}
{Sharma}, P., {McCourt}, M., {Parrish}, I.~J.,  \& {Quataert}, E. 2012, \mnras,
  427, 1219

\bibitem[\protect\citeauthoryear{{Sharma} et~al.}{{Sharma}
  et~al.}{2014}]{2014MNRAS.443.3463S}
{Sharma}, P., {Roy}, A., {Nath}, B.~B.,  \& {Shchekinov}, Y. 2014, \mnras, 443,
  3463

\bibitem[\protect\citeauthoryear{{Shull} et~al.}{{Shull}
  et~al.}{2015}]{2015ApJ...811....3S}
{Shull}, J.~M., {Moloney}, J., {Danforth}, C.~W.,  \& {Tilton}, E.~M. 2015,
  \apj, 811, 3

\bibitem[\protect\citeauthoryear{{Socrates}, {Davis}, \&
  {Ramirez-Ruiz}}{{Socrates} et~al.}{2008}]{2008ApJ...687..202S}
{Socrates}, A., {Davis}, S.~W.,  \& {Ramirez-Ruiz}, E. 2008, \apj, 687, 202

\bibitem[\protect\citeauthoryear{{Spitzer}}{{Spitzer}}{1965}]{spitzer65}
{Spitzer}, L. 1965, {Physics of fully ionized gases}

\bibitem[\protect\citeauthoryear{{Stanimirovi{\'c}} et~al.}{{Stanimirovi{\'c}}
  et~al.}{2002}]{stanimirovic02}
{Stanimirovi{\'c}}, S., {Dickey}, J.~M., {Kr{\v c}o}, M.,  \& {Brooks}, A.~M.
  2002, \apj, 576, 773

\bibitem[\protect\citeauthoryear{{Stern} et~al.}{{Stern}
  et~al.}{2016}]{stern16}
{Stern}, J., {Hennawi}, J.~F., {Prochaska}, J.~X.,  \& {Werk}, J.~K. 2016,
  \apj, 830, 87

\bibitem[\protect\citeauthoryear{{Stinson} et~al.}{{Stinson}
  et~al.}{2012}]{stinson12}
{Stinson}, G.~S., et~al. 2012, \mnras, 425, 1270

\bibitem[\protect\citeauthoryear{{Stocke} et~al.}{{Stocke}
  et~al.}{2013}]{stocke13}
{Stocke}, J.~T., {Keeney}, B.~A., {Danforth}, C.~W., {Shull}, J.~M., {Froning},
  C.~S., {Green}, J.~C., {Penton}, S.~V.,  \& {Savage}, B.~D. 2013, \apj, 763,
  148

\bibitem[\protect\citeauthoryear{{Suresh} et~al.}{{Suresh}
  et~al.}{2015}]{suresh15}
{Suresh}, J., {Rubin}, K.~H.~R., {Kannan}, R., {Werk}, J.~K., {Hernquist}, L.,
  \& {Vogelsberger}, M. 2015, ArXiv:1511.00687

\bibitem[\protect\citeauthoryear{{Thompson} et~al.}{{Thompson}
  et~al.}{2016}]{thompson16}
{Thompson}, T.~A., {Quataert}, E., {Zhang}, D.,  \& {Weinberg}, D.~H. 2016,
  \mnras, 455, 1830

\bibitem[\protect\citeauthoryear{{Thornton} et~al.}{{Thornton}
  et~al.}{1998}]{1998ApJ...500...95T}
{Thornton}, K., {Gaudlitz}, M., {Janka}, H.-T.,  \& {Steinmetz}, M. 1998, \apj,
  500, 95

\bibitem[\protect\citeauthoryear{{Tumlinson} et~al.}{{Tumlinson}
  et~al.}{2013}]{tumlinsonCOS}
{Tumlinson}, J., et~al. 2013, \apj, 777, 59

\bibitem[\protect\citeauthoryear{{Tumlinson} et~al.}{{Tumlinson}
  et~al.}{2011}]{tumlinson11}
{Tumlinson}, J., et~al. 2011, Science, 334, 948

\bibitem[\protect\citeauthoryear{{Upton Sanderbeck} et~al.}{{Upton Sanderbeck}
  et~al.}{2017}]{uptonsanderbeck}
{Upton Sanderbeck}, P.~R., {McQuinn}, M., {D'Aloisio}, A.,  \& {Werk}, J.~K.
  2017, ArXiv:1710.07295

\bibitem[\protect\citeauthoryear{{Wakker} \& {van Woerden}}{{Wakker} \& {van
  Woerden}}{1997}]{1997ARA&A..35..217W}
{Wakker}, B.~P.,  \& {van Woerden}, H. 1997, \araa, 35, 217

\bibitem[\protect\citeauthoryear{{Werk} et~al.}{{Werk} et~al.}{2016}]{werk16}
{Werk}, J.~K., et~al. 2016, \apj, 833, 54

\bibitem[\protect\citeauthoryear{{Werk} et~al.}{{Werk} et~al.}{2013}]{werk13}
{Werk}, J.~K., {Prochaska}, J.~X., {Thom}, C., {Tumlinson}, J., {Tripp}, T.~M.,
  {O'Meara}, J.~M.,  \& {Peeples}, M.~S. 2013, \apjs, 204, 17

\bibitem[\protect\citeauthoryear{{Werk} et~al.}{{Werk} et~al.}{2014}]{werk14}
{Werk}, J.~K., et~al. 2014, \apj, 792, 8

\bibitem[\protect\citeauthoryear{{Wiersma}, {Schaye}, \& {Smith}}{{Wiersma}
  et~al.}{2009}]{2009MNRAS.393...99W}
{Wiersma}, R.~P.~C., {Schaye}, J.,  \& {Smith}, B.~D. 2009, \mnras, 393, 99

\bibitem[\protect\citeauthoryear{{Zhang} et~al.}{{Zhang}
  et~al.}{2014}]{zhang14}
{Zhang}, D., {Thompson}, T.~A., {Murray}, N.,  \& {Quataert}, E. 2014, \apj,
  784, 93

\bibitem[\protect\citeauthoryear{{Zheng} et~al.}{{Zheng}
  et~al.}{2015}]{2015ApJ...807..103Z}
{Zheng}, Y., {Putman}, M.~E., {Peek}, J.~E.~G.,  \& {Joung}, M.~R. 2015, \apj,
  807, 103

\end{thebibliography}

\appendix

\section{A. survival times and velocities of clouds}
\label{sec:kinematics}

As $T\lesssim T_{\rm vir}$K gas ``clouds'' are created, such as from thermal instability, they are likely to lose buoyancy and fall in the potential well.  After falling for a dynamical time from $\sim r_{\rm vir}$, or $\sim 1~$Gyr, such clouds will reach velocities of hundreds of kilometers per second if hydrodynamic instabilities do not destroy them and if gaseous drag is negligible.  The observed ``clouds'' in the CGM show a large range of line-of-sight velocities, with some exceeding the circular velocity but few with velocities larger than the escape velocity \citep{werk16}.  We argued in the main text that the coherence of absorbers -- that in the same system clouds appear to be highly clustered -- suggests that the virialized gas is sloshing with typical projected velocities of $50-150~$km~s$^{-1}$ and that much of the kinematics does not owe to independent motions of individual clouds.  In the sloshing picture, the warm and cold gases are entrained in the virialized atmosphere.   Here we estimate the timescales for the survival and entrainment of clouds.

  Gaseous drag decelerates a cloud of mass $M_{\rm cloud}$ and cross-sectional area $A$ with force of $-1/2 \rho  v^2 A C_D$, where $\rho$ is the mass densities of the cloud and background, $C_D$ is the drag coefficient (that is likely of the order of unity):
  \begin{eqnarray}
v_{\rm term} &=& \sqrt{\frac{2 M_{\rm cloud}' g}{C \rho A}} = 100 {\rm~ km ~s}^{-1} C_D^{-1/2}\left( \frac{\eta-1}{10}  \frac{R_{\rm cloud}}{1~{\rm kpc}} \right)^{1/2} \left(\frac{g}{(200 {\rm \, km\, s}^{-1})^2/100 {\rm \, kpc}}\right)^{1/2},
\end{eqnarray}
where we have approximated the clouds as spheres of radius $R$, $M_{\rm cloud}'$ subtracts out the mass in the background gas from $M_{\rm cloud}$ to account for buoyancy,  and $\eta \equiv n/n_{\rm bk}$ where $n_{\rm bk}$ is the coronal number density.  We expect  $\eta \sim 3$ for warm \OVI-tracing gas and $\sim 30-100$ for thermal pressure-supported photoionized gas (although we find evidence that the cold photoionized gas is nonthermally supported).  For reference, a $z=0.2$ NFW halo with $M_{\rm halo}=10^{12}\Msun$ and concentration parameter of $12$,  has $g =1200$, $400$, and 150 (km s$^{-1}$)$^2$ kpc$^{-1}$ at radii of $50, 100$ and $200\;$kpc. The timescale to reach the terminal velocity is
\begin{equation}
t_{\rm term} \equiv \frac{v_{\rm term}}{g} = 250 ~{\rm Myr} \, C_D^{-1/2} \left(  \frac{\eta-1}{10} \frac{R_{\rm cloud}}{1~{\rm kpc}} \right)^{1/2} \left(\frac{g}{(200 {\rm \, km\, s}^{-1})^2/100 {\rm \, kpc}}\right)^{-1/2}.
\end{equation}
Similarly, if we consider a cloud moving at some velocity, the timescale for it to become decelerated by gaseous drag is similar:
\begin{eqnarray}
t_{\rm stop} &\equiv & \frac{M_{\rm cloud} v}{F_{\rm drag}}  = 260 \, {\rm Myr} ~C_D^{-1}~ v_{2}^{-1} \left(\frac{\eta}{10} \frac{R_{\rm cloud}}{1~{\rm kpc}}\right).
\end{eqnarray}
On the other hand, one would expect the cloud is crushed on the timescale for a shock to cross the cloud or 
\begin{equation}
t_{\rm crush} \equiv \left(\frac{n_{\rm cloud}}{n_{\rm bk}} \right)^{1/2} \frac{R_{\rm cloud}}{v} = 30 {\rm ~Myr} \left(\frac{\eta}{10} \right)^{1/2} \left( \frac{R_{\rm cloud}}{1~{\rm kpc}} \right) v_{2}^{-1}.
\end{equation}
Other cloud destruction mechanisms, namely Kelvin Helmholtz and Rayleigh Taylor instabilities, operate on similar timescales to $t_{\rm crush}$ (although, for both, magnetic fields and cooling can reduce the rate of destruction).  The shortness of $t_{\rm crush}$ relative to the dynamical time suggests that thermally unstable droplets may not survive long enough to accelerate to hundreds of kilometers per second, perhaps explaining the apparent entrainment of warm and cold clouds.  Nonthermal pressure support, which reduces $\eta$, results in clouds that are more entrained by gaseous drag, but also that are more susceptible to hydrodynamic instabilities.  

\section{B. Proximity radiation and self-photoionization}
\label{sec:proximity}
 An enhancement in the ionizing background at the ionization potential for \OVI\ ($138\;$eV) over standard ionizing background models would result in higher densities bearing more \OVI\ for $10^4$K gas \citep{werk16}.  A factor of $\approx 30-300$ larger densities would be required for photoionized \OVI\ models to be consistent with Galactic observations that suggest $P= 10-100~$K~cm$^{-3}$ and hence $n= 10^{-3}-10^{-2}~$cm$^{-2}$, assuming thermal pressure support. This appendix argues that ionizing radiation background models such as by \citet{haardt12} cannot so drastically underestimate the true $z=0$ background.
 
First, a large enhancement cannot owe to proximity radiation from the local galaxy.  For example, if galactic ionizing emission traces star formation, the radius at which the emission from the local galaxy equals the background is
\begin{equation}
 r_{\rm prox} = 90 ~~{\rm kpc} ~~ \sqrt{\frac{{\rm SFR}}{1 ~\Msun {\rm~yr}^{-1} }},
 \label{eqn:rprox}
\end{equation}
where SFR denotes the galaxy's star formation rate.  This relation is calculated at $z=0$ assuming a quasar-like spectral index in specific luminosity of $\alpha= 1.5$ for the emission ($ r_{\rm prox}  = 130~{\rm kpc}$ if instead $\alpha=2.5$) and an empirically motivated $\dot \rho_{\rm SFR} = 0.02 (1+z)^{2} \Msun {\rm comoving~ Mpc}^{-3}$yr$^{-1}$ for $z\lesssim 2$ \citep{hopkinsbeacom}.  We remark that $r_{\rm prox}(z=0)$ is insensitive to $z>2$ and that $r_{\rm prox}$ decreases with increasing redshift for fixed SFR.  Equation~(\ref{eqn:rprox}) conservatively assumes that the unspecified emission process, rather than the canonical source active galactic nuclei, AGNs, dominates the ionizing background.  [Equation~(\ref{eqn:rprox}) is calculated using that the average specific intensity is 
\begin{equation}
 J_{\nu_0}(z=0) = \frac{1}{4 \pi} \int_0^\infty \frac{c \, dz}{H(z) (1+z)}  \zeta_{\nu} \, \dot  \rho_{\rm SFR}(z),
 \end{equation}
  where $\nu = \nu_0 (1+z)$, $\dot  \rho_{\rm SFR}(z)$ is again the comoving rate density, and we have used that the distance traveled by $> 20$\,eV photons radiated at $z<2$ is, for most photons, limited by light travel.  We then equate $J_{\nu_0}$ to 
 \begin{equation} 
  I_{\nu_0}^{\rm prox} = \frac{\rm \zeta_{\nu_0}  \, SFR}{4 \pi r^2},
  \end{equation}
   where $\zeta_\nu$ converts a star formation rate to a specific luminosity, and solve for $r$.]  In conclusion, since \OVI\ absorbers extend to the virial radius, most of these absorbers cannot have more than a modest enhancement from local radiation.

This conclusion also applies to cases where there is a burst of ionizing photons, such as if a `flaring' AGN in the host galaxy, that turns on every $t_{\rm AGN}$ and flash ionizes the CGM \citep{2013MNRAS.434.1063O}.  If sufficiently intense, this flash could result in the ionization of select metals being vastly out of equilibrium with the metagalactic background.  However, when the ambient radiation is averaged over a time $t_{\rm AGN}$, the more temporally constant ionizing background from all such sources is comparable to the amount of bursty flux at a radius given by equation~\ref{eqn:rprox} (noting that this bursty AGN activity may not trace star formation and so $r_{\rm prox}$ could differ from the estimate in eqn.~\ref{eqn:rprox} at the ${\cal O}(1)$ level).  {Thus, such flash over-ionization can only affect the $\gtrsim 100\,$kpc CGM by more than an ${\cal O}(1)$ factor on a timescale $t \ll t_{\rm AGN}$, with the relative contribution of the flash decreasing with radius as $r^{-2}$.  For the majority of the time, the CGM ionization will be set by the ionizing background and, hence, the bulk of the \OVI, which resides at $\gtrsim 100~$kpc, cannot be generated by out-of-equilibrium oxygen from local bursts of ionizing flux.  This qualitative argument is not inconsistent with the results of \citet{2017arXiv170507897O}, who find that some models for flaring AGNs can yield a factor of $2-3$ enhancement at $R<140~$kpc.}

While the above arguments show that a large enhancement cannot owe to proximity effects, it is possible that standard ionizing background models such as by \citet{haardt01} vastly underestimate the true ionizing background.  However, there are constraints on where this enhancement can happen, as at $\approx 13.6\;$eV and $\gtrsim 300\;$eV standard $z=0$ ionizing backgrounds are consistent at the factor of $\sim 2$ level with the background inferred from the Ly$\alpha$ forest \citep{2015ApJ...811....3S,2015MNRAS.451L..30K} and soft X-ray background measurements \citep{haardt12}.  Indeed the background relevant for high ions is constrained better at $z=0$ than at higher redshifts because of the X-ray background measurements.  While the ionization potential for \OVI\ ( $138\;$eV) falls far from the energies constrained by these observational diagnostics, it is difficult to envision more than a factor of several underestimate from the physically motivated interpolation performed by ionizing background models.  See \citet{uptonsanderbeck} for more discussion.

Finally, perhaps self-photoionization associated with the cooling gas itself changes the ionization state of CGM clouds.  Self-photoionization is often important in shock and cooling flow models \citep{1991PASP..103..923S, 2009ApJ...693.1514G}.  However, the extremely low densities of the CGM should make such radiation less important there.  Indeed, if the emitting region is comparable to $r_{\rm vir}$, the proximity region arguments are even more constraining for the importance of self-photoionization:  A ray originating at $z\sim 0$ and extending to to $z=1$ intersects several $M_{\rm halo} >10^{12} \Msun$ halos within $1 \,r_{\rm vir}$, with this number increasing quickly with decreasing halo mass threshold \citep{2014ApJ...780L..33M}.  By conservation of intensity (and ignoring redshifting), each CGM contributes equally to the brightness as that contributed locally.  Thus, the local emission of ionizing flux by cooling gas at large $R$ is unlikely to dominate over the background sourced by the same emission process.  This argument further implies that ionizing emission emitted by the CGM is constrained by ultraviolet and soft X-ray background measurements \citep{uptonsanderbeck}. 

\section{C. Cooling times and ionized fraction}
\label{sec:helpful}

\begin{figure}
\begin{center}
\epsfig{file=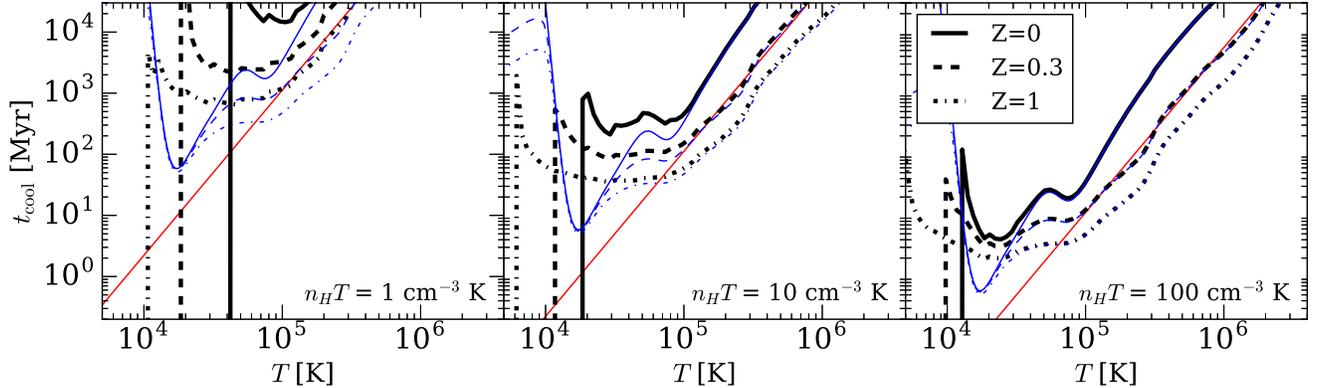, width=18cm}
\end{center}
\caption{Isobaric gas cooling times assuming the \citet{haardt01} photoionizing background (thick black curves) or no background (thin blue curves) at the specified metallicities.  These calculations do not assume ionization equilibrium and instead use the non-equilibrium cooling rates for isobaric cooling from a high temperature, although we find that it makes little difference if instead equilibrium is assumed.  The thin red line is the cooling time given by equation~(\ref{eqn:tcool}) for $Z=0.3$. \label{fig:tcool}}
\end{figure}

\begin{figure*}
\begin{center}
\epsfig{file= 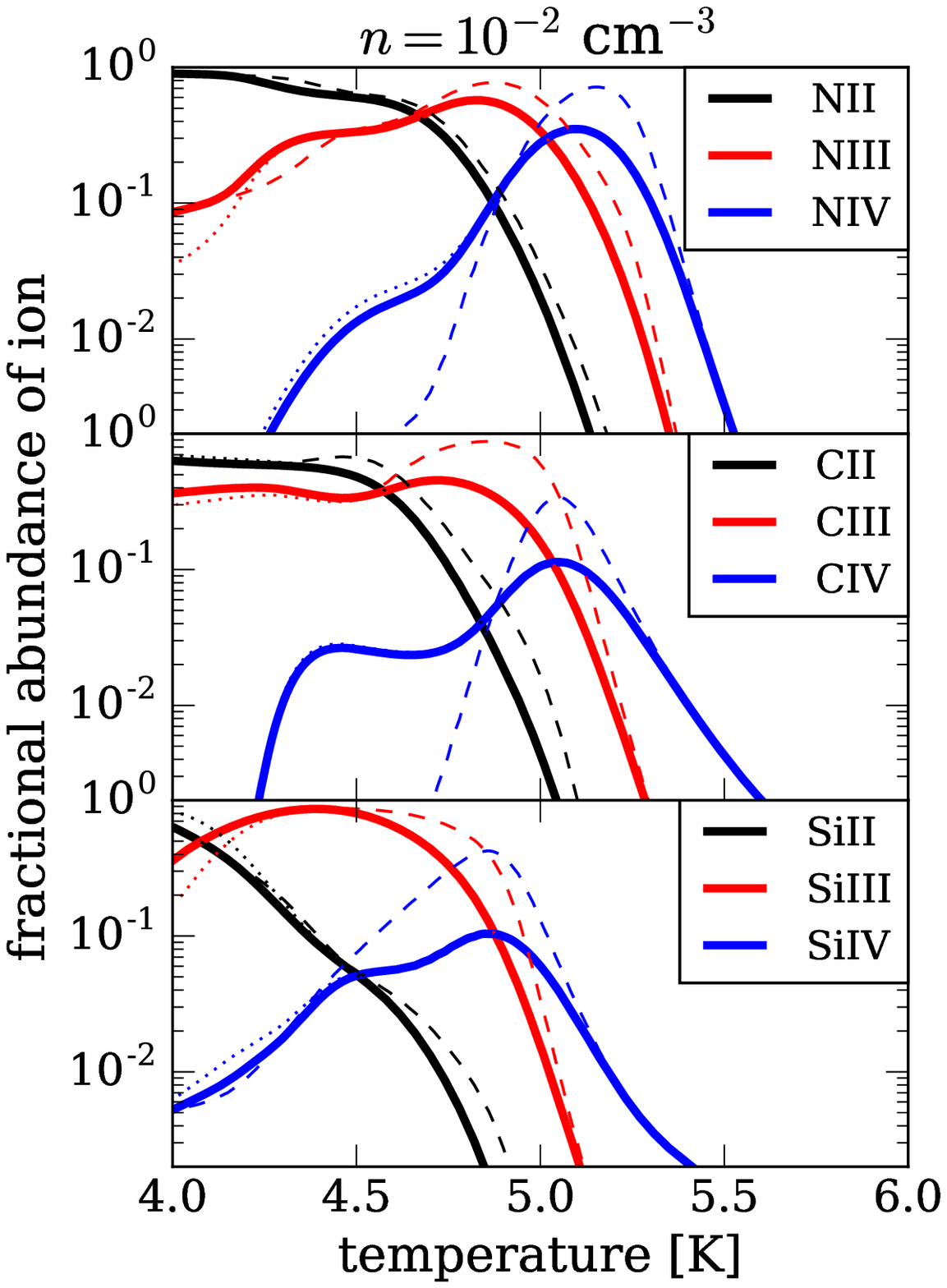, width=5.5cm}
\epsfig{file= 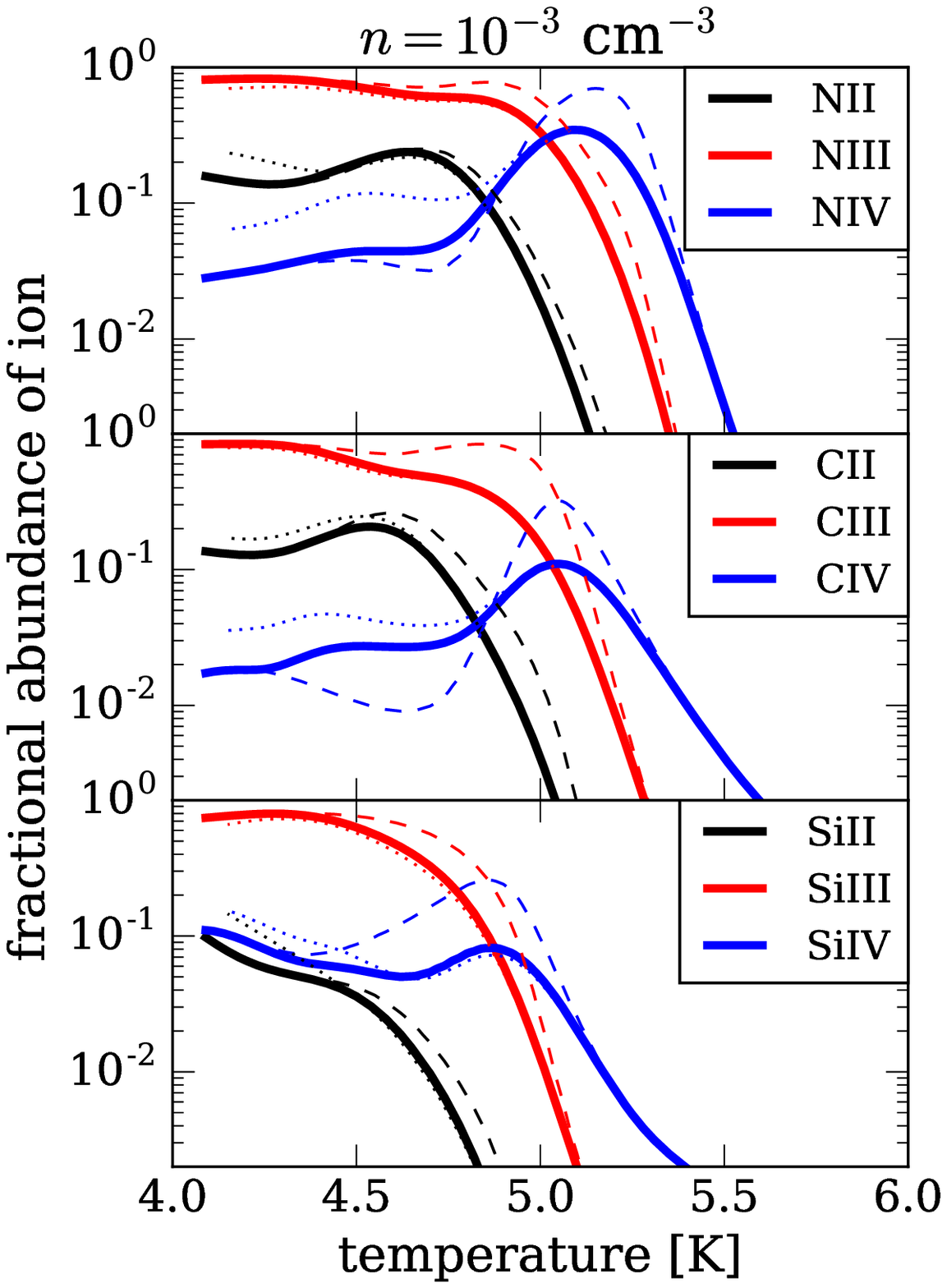, width=5.5cm}
\epsfig{file= 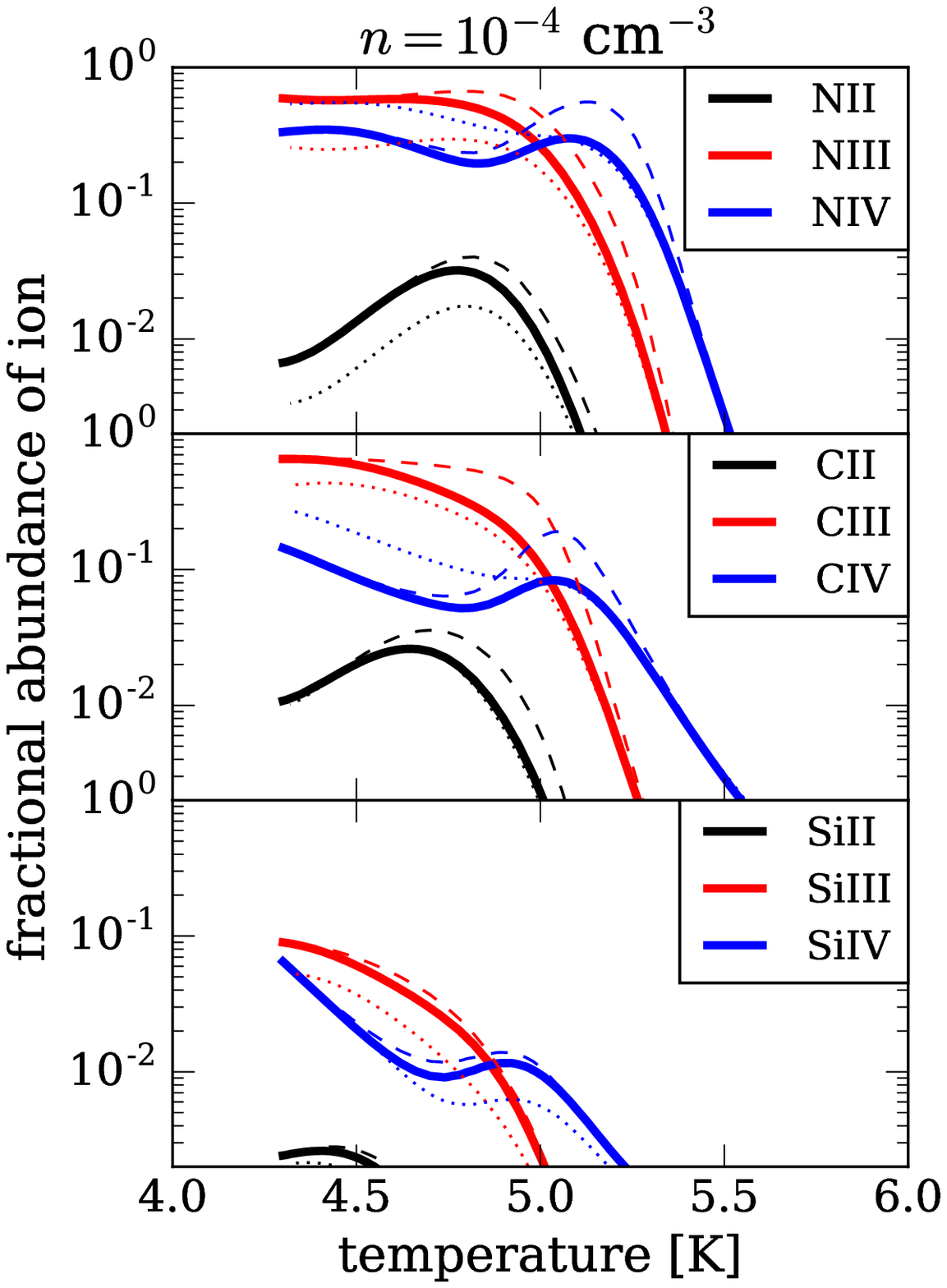, width=5.5cm}
\end{center}
\caption{The solid curves show the ionic fraction for the fiducial \citet{haardt01} background assuming $Z=0.3$ and isochoric cooling from $10^7~$K at the density specified at the top  using the non-equilibrium calculations of \citet{oppenheimer13}.  The dotted and dashed curves show the same except, respectively, for the \citealt{haardt12} ultraviolet background and assuming ionization equilibrium.  The lowest temperatures at which these curves terminate mark thermal equilibrium.  In Section~\ref{sec:HI}, we use that there are {\cal O}(1) fractions in the twice-ionized species of these metals for $n\sim 10^{-3}$cm$^{-3}$ and $T=10^4-10^5\;$K to constrain the densities of the photoionized CGM clouds. \label{fig:Xion} }
\end{figure*}

The text often refers to cooling times and metal ionization fractions.  Figure~\ref{fig:tcool} shows the cooling times for different metallicities, and Figure~\ref{fig:Xion} shows the ionic fractions for nitrogen, carbon, and silicon as a function of temperature (which are pertinent to the discussion in Section~\ref{sec:HI}).  The curves of different line styles illustrate the effects of different ionizing backgrounds and different assumptions regarding ionization equilibrium, and the thin red line is the simple parametric form for the cooling time used in our analytic calculations (eqn.~\ref{eqn:tcool}) for $Z=0.3$.  See the figure captions for additional details.

\bibliographystyle{apj}

\end{document}